\documentclass[a4paper,11pt]{article}
\pdfoutput=1 

\usepackage{jinstpub} 
\usepackage[]{units}
\usepackage[colorinlistoftodos]{todonotes}
\DeclareMathOperator\erf{erf}
\usepackage{subcaption}
\usepackage{tikz}
\usetikzlibrary{spy}
\usetikzlibrary{calc}
\def\centerarc[#1](#2)(#3:#4:#5)
    { \draw[#1] ($(#2)+({#5*cos(#3)},{#5*sin(#3)})$) arc (#3:#4:#5); }
\usetikzlibrary{decorations.pathmorphing}
\usetikzlibrary{decorations.pathreplacing}

\usetikzlibrary{shapes,backgrounds}
\newcommand{\tstar}[5]{
\pgfmathsetmacro{\starangle}{360/#3}
\draw[#5] (#4:#1)
\foreach \x in {1,...,#3}
{ -- (#4+\x*\starangle-\starangle/2:#2) -- (#4+\x*\starangle:#1)
}
-- cycle;
}

\usepackage{tabularx}
\usepackage{multirow}
\newcommand{\ra}[1]{\renewcommand{\arraystretch}{#1}}
\usepackage{booktabs}

\title{\boldmath Muon reconstruction with a geometrical model in JUNO}

\author[a,b,1]{C. Genster\note{Corresponding author.}}
\author[a,b]{M. Schever}
\author[a,b]{L. Ludhova}
\author[b]{M. Soiron}
\author[b]{A. Stahl}
\author[b]{C. Wiebusch}
\affiliation[a]{Forschungszentrum J\"ulich IKP,\\Wilhelm-Johnen-Strasse, D-52428 J\"ulich, Germany}
\affiliation[b]{III. Physikalisches Institut B, RWTH Aachen University, Aachen, Germany}

\emailAdd{c.genster@fz-juelich.de}

\abstract{
The Jiangmen Neutrino Underground Observatory (JUNO) is a 20\,kton liquid scintillator detector currently under construction near Kaiping in China. The physics program focuses on the determination of the neutrino mass hierarchy with reactor anti-neutrinos. For this purpose, JUNO is located \unit[650]{m} underground with a distance of \unit[53]{km} to two nuclear power plants. As a result, it is exposed to a muon flux that requires a precise muon reconstruction to make a veto of cosmogenic backgrounds viable.
Established muon tracking algorithms use time residuals to a track hypothesis. We developed an alternative muon tracking algorithm that utilizes the geometrical shape of the fastest light. It models the full shape of the first, direct light produced along the muon track. From the intersection with the spherical PMT array, the track parameters are extracted with a likelihood fit. The algorithm finds a selection of PMTs based on their first hit times and charges. Subsequently, it fits on timing information only. On a sample of through-going muons with a full simulation of readout electronics, we report a spatial resolution of \unit[20]{cm} of distance from the detector's center and an angular resolution of $\unit[1.6]{^{\circ}}$ over the whole detector.
Additionally, a dead time estimation is performed to measure the impact of the muon veto. Including the step of waveform reconstruction on top of the track reconstruction, a loss in exposure of only 4\% can be achieved compared to the case of a perfect tracking algorithm. When including only the PMT time resolution, but no further electronics simulation and waveform reconstruction, the exposure loss is only 1\%.
}


\begin{document}
\maketitle
\flushbottom

\section{Introduction}
\label{sec:intro}

The Jiangmen Underground Neutrino Observatory (JUNO) is a reactor anti-neutrino experiment currently being built close to Kaiping, China. Its main goal is the determination of the neutrino mass hierarchy with the inverse beta decay (IBD) $ \bar{\nu}_e + p \rightarrow n + e^+$. The broad physics program \cite{yb:2015} and the unprecedented size dictate the detector design described in detail in \cite{cdr:2015}.
For maximum sensitivity to the neutrino mass hierarchy, the experimental site of JUNO is confined to a distance of \unit[53]{km} from the two nuclear power plants (NPP), Yangjiang NPP and Taishan NPP. The underground laboratory will have \unit[650]{m} rock overburden, resulting in a muon rate in the central detector (CD) of $\unit[3.5]{s^{-1}}$ with a mean energy of \unit[215]{GeV} \cite{yb:2015}. Apart from overshadowing any other event while moving through the detector, muons will produce cosmogenic isotopes. $^9$Li and $^8$He are the most prominent cosmogenic backgorunds. They have a long lifetime of \unit[256]{ms} and \unit[172]{ms} \cite{TILLEY2004155}, respectively and their $(\beta^-+ n)$ decay-channel mimics the correlated decay of IBD events. A detailed study of the production characteristics \cite{Grassi:2014hxa,Grassi:2015fua} showed that 99\% of the cosmogenics are within \unit[3]{m} distance to the muon track following approximately an exponential distribution. Therefore, it is possible to veto the cosmogenic background with a cylindrical volume around the muon track. In order to achieve a high efficiency with this method, the track reconstruction of muons and position reconstruction of $^9$Li and $^8$He decays need sufficiently good resolutions. A cylindrical veto with \unit[3]{m} radius after a single muon track will take less than 5\% of the target volume, due to the unprecedented size of JUNO.

\begin{figure}[htbp]
\centering 
\includegraphics[width=\textwidth]{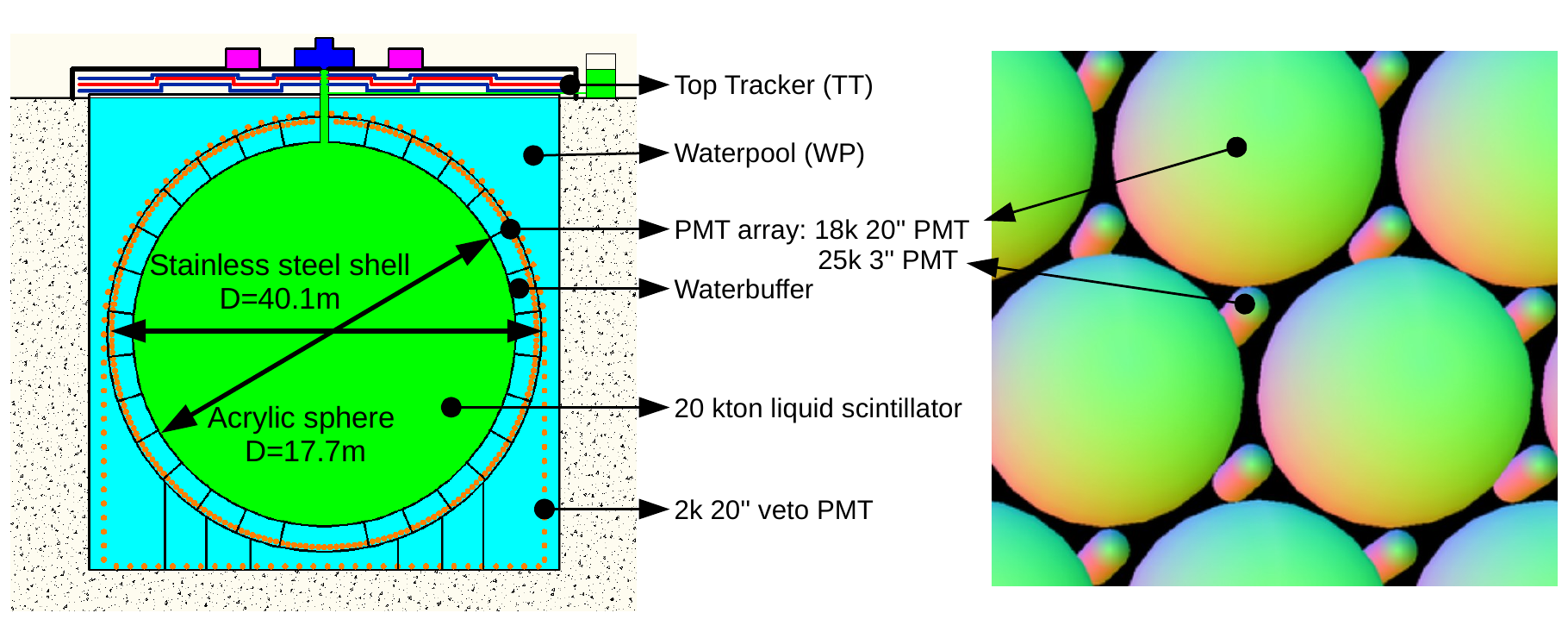}
\caption{\label{fig:detschematic} Schematic cross section of the detector. The acrylic sphere holding the liquid scintillator together with the arrays of large and small PMTs and the intermediate water buffer are referred to as the Central Detector. On the right, the placement of 3-inch PMTs between the 20-in PMTs is shown.}
\end{figure}

The innermost part of JUNO is the Central Detector (CD), which is placed in a cylindrical waterpool (WP). The CD consists of an acrylic sphere of $D_{LS} = 2R_{LS} = \unit[35.4]{m}$ diameter that holds \unit[20]{kton} of LAB-based liquid scintillator. It is retained by a stainless steel shell, which also holds 18000 20-inch and 25000 3-inch photomultiplier tubes (PMTs) at a radius $R_{CD} =\unit[19.5]{m}$. As can be seen in figure \ref{fig:detschematic}, a waterbuffer of about \unit[1.7]{m} is located between the acrylic sphere and the PMT array. The liquid scintillator contains a mixture of PPO as a flour and the wavelength-shifter Bis-MSB to obtain the optimum balance between the light yield and attenuation lenght \cite{cdr:2015}. The scintillator's light yield is 1200 photo electrons (p.e) per MeV deposited energy. This results in an energy resolution of $3\%/\sqrt{E(\rm{MeV})}$ for point-like events. Muons create a signal in the order of $10^7$ p.e. per event.
The WP that surrounds the CD has a diameter of \unit[43.5]{m} and a height of \unit[44]{m}. It is filled with ultra-pure water and optically separated from the CD. Instrumented with 2000 20-inch PMTs, it acts as a water Cherenkov detector to aid the muon tracking effort. Additionally, it also shields the CD from external radioactivity.
On top of the waterpool, the Top Tracker (TT) is placed. It is made up of the former OPERA target tracker \cite{Adam:2007ex}. The plastic scintillator walls can only cover the central segment above the WP, but they can provide a calibration set of high precision track-starting points.
The CD of JUNO features two independent PMT systems. The 20-inch PMT array (LPMT) covers about 75\% of the spherical area and consists of two different kinds of PMTs. There are about 5000 dynode PMTs from Hamamatsu with a transit time spread (TTS) of \unit[3]{ns}. The other 13000 PMTs are multi-channel-plate (MCP)-PMTs developed at IHEP, China with a TTS of \unit[12]{ns}. The larger TTS originates from the different architecture of the MCP-PMTs. The 3-inch PMTs (SPMTs) will be installed in the spaces between the LPMTs (right panel in figure~\ref{fig:detschematic}) and act as a complementary system. This PMT array accounts for 2.5\% coverage and collect much less light due to that. In return, they have a smaller dynamic range and are not prone to saturation in case of a muon event. The SPMTs are also fast with a TTS of less than \unit[5]{ns}.
The readout electronics of both systems are submerged in the waterpool in order to quickly convert and process the analogue PMT signal.

\section{Characteristics of muon events}
\label{sec:char}
\begin{figure}[htbp]
\centering
	\begin{tikzpicture}[scale=1.8, spy using outlines]
	\coordinate (center) at (0,0);
	\coordinate (startleft) at (-0.4,2.2);
	\coordinate (endleft) at (-0.4,-2.2);
	\coordinate (startright) at (1.7,2.2);
	\coordinate (endright) at (1.7, -2.2);
	
	\coordinate (inCDleft) at (-0.4,1.908533);
	\coordinate (outCDleft) at (-0.4,-1.908533);
	\coordinate (inLSleft) at (-0.4,1.724210);
	\coordinate (outLSleft) at (-0.4,-1.724210);
	
	\coordinate (inCDright) at (1.7,0.955249);
	\coordinate (outCDright) at (1.7,-0.955249);
	\coordinate (inLSright) at (1.7,0.492849);
	\coordinate (outLSright) at (1.7,-0.492849);
	
	\draw (center) circle [radius=1.77]; 
	\draw (center) circle [radius=1.95]; 
	
	\draw [thick] (startleft) node [scale=0.75,left] {$track_1$} -- (inCDleft);
	\draw [thick,->] (outCDleft) -- (endleft);
	\draw [black] (center) -- (-0.4,0); \node [scale=0.65] at (-0.2,-0.15) {$D_1$};
	\centerarc[black](-0.4, 0)(0:90:0.15)
	\draw [fill] (-0.335, 0.065) circle [radius=0.01];
	\draw [orange, thick] (center) -- (inCDleft) node [scale=0.5,midway, above, sloped] {$R_{CD}$};
	\draw [purple, thick] (center) -- (outLSleft) node [scale=0.5,midway, above, sloped, xshift=15pt] {$R_{LS}$};
	\draw [thick, blue] (inCDleft) -- (inLSleft);
	\draw [thick, blue] (outLSleft) -- (outCDleft);
	\draw [thick, black] (inLSleft) -- (outLSleft);
	\draw [thick, green, dashed] (inLSleft) -- (outLSleft);

	\draw [thick] (startright) node [scale=0.75,left] {$track_2$} -- (inCDright);
	\draw [thick,->] (outCDright) -- (endright);
	\draw [black] (center) -- (1.7,0); \node [scale=0.65] at (0.85,0.12) {$D_2$};
	\centerarc[black](1.7, 0)(90:180:0.15)
	\draw [fill] (1.7-0.15/2.+0.02, 0.065) circle [radius=0.01];
	\draw [orange, thick] (center) -- (inCDright) node [scale=0.5,midway, above, sloped] {$R_{CD}$};
	\draw [purple, thick] (center) -- (outLSright) node [scale=0.5,midway, above, sloped, xshift=15pt] {$R_{LS}$};
	\draw [thick, blue] (inCDright) -- (inLSright);
	\draw [thick, blue] (outLSright) -- (outCDright);
	\draw [thick, black] (inLSright) -- (outLSright);
	\draw [thick, green, dashed] (inLSright) -- (outLSright);
	
	\draw [fill] (inCDright) circle [radius=0.025];
	\draw [fill] (outLSright) circle [radius=0.025];
	\draw [fill] (center) circle [radius=0.025];
	\draw [fill] (inCDleft) circle [radius=0.025];
	\draw [fill] (outLSleft) circle [radius=0.025];
	
	\spy [red, thick, draw, height=1.5*5.5cm,width=1.5*2cm,magnification=2, connect spies] on (1.5*1.2*1.7,0) in node [fill=white] at (1.2*3.5, 0);
		
	\end{tikzpicture}
	\caption{\label{fig:trackLengths} Track lengths in different media in the JUNO CD. Track 1 has a distance from center $D_1 = \unit[4]{m}$, while track 2 is closer to the edge with $D_2=\unit[17]{m}$. For both tracks the length in the waterbuffer is marked in blue and the track length in LS is given by the green dashed line. The inset gives a magnified view on the track close to edge to show that almost equal lengths in LS and waterbuffer are traversed.}
\end{figure}
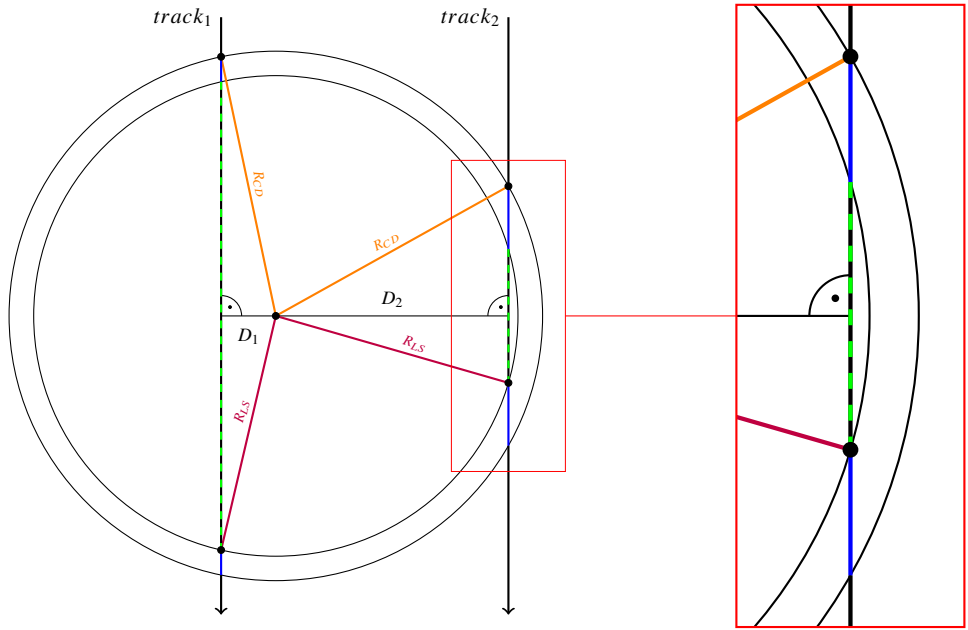

\begin{figure}[htbp]
\centering
\includegraphics[width=.75\textwidth]{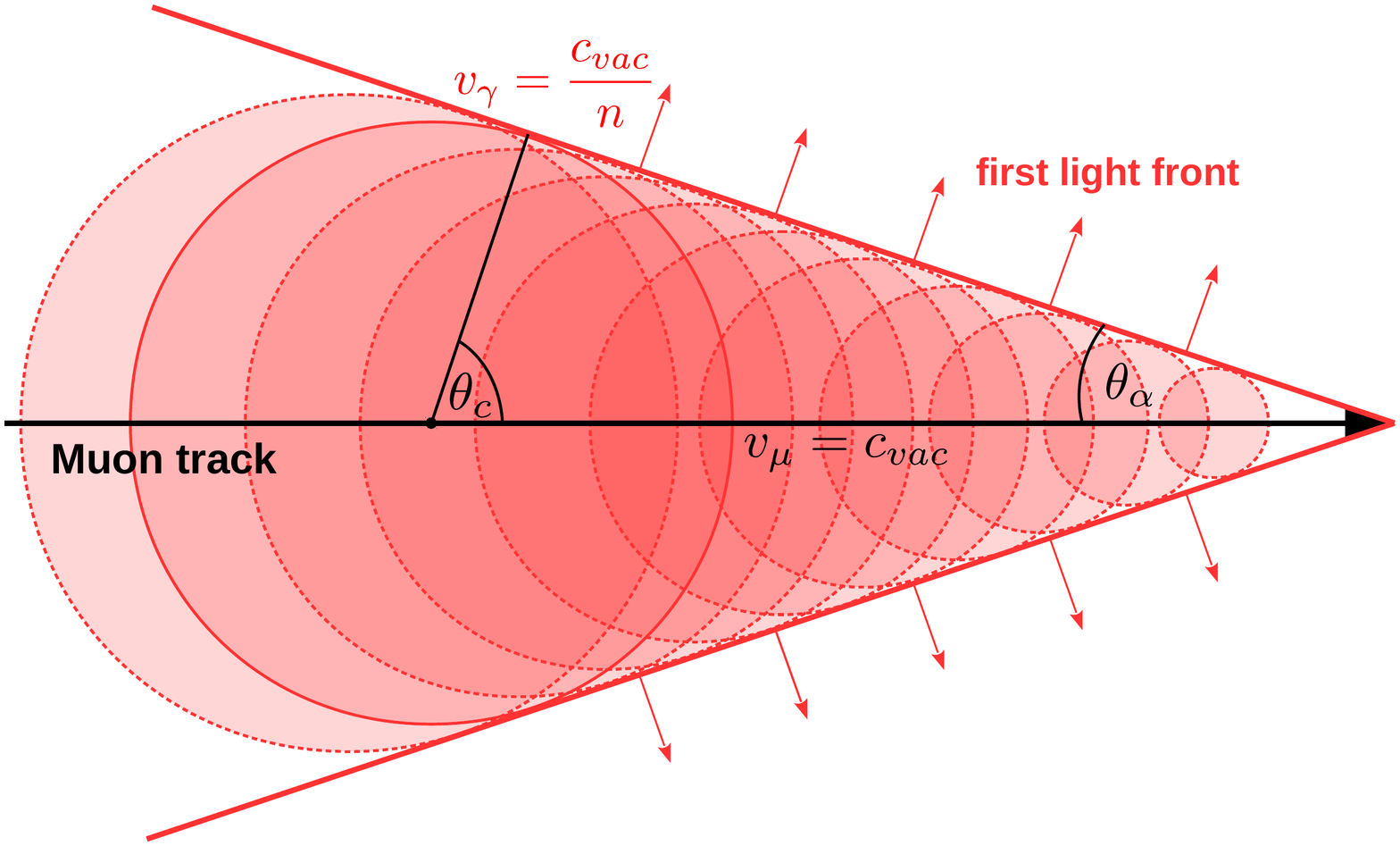}
\caption{\label{fig:construction} Build up of the first-light surface by isotropic emission of photons along a muon track in liquid scintillator. The opening angle $\theta_\alpha$ depends on the photons group velocity and by that on the refractive index $n$ of the traversed medium.}
\end{figure}

Muon tracks are the brightest events expected in JUNO, which gives them a distinct signature. Nevertheless, the detailed characteristics depend mainly on the muon's tracklength in liquid scintillator and in water. Before reaching the CD, a muon will travel through a part of the cylindrical waterpool and create Cherenkov light. Due to the optical seperation of the WP from the CD, this light will only be detected by the waterpool PMTs. Whether or not it crosses the CD can then be distinguished by the CD signal. Given the energy loss only by ionization of \unit[1.43]{MeV/cm} \cite{Grassi:2014hxa}, even a corner-clipping muon with \unit[1]{m} track length in LS would produce more than $\unit[1.5\cdot10^5]{p.e}$. This amount of light is sufficient to tag muons.
For the purpose of background rejection, the muons travelling through the CD are the most important. When entering the CD, the muon will first traverse the waterbuffer between the PMT array and the acrylic sphere. On this path, it will create Cherenkov light just like in the waterpool. After that, the acrylic sphere is entered and the LS is crossed. When exiting the detector, the muon will traverse the waterbuffer once more. Since the light production in water and LS is very different, the tracklength in each medium has a large impact on event characteristics. Under the assumption of a straight line track and with a given track's minimal distance from the detector's center $D$, the tracklength in the CD can be expressed as $l_{\rm{CD}} = 2\sqrt{R^2_{\rm{CD}} - D^2}$. With the tracklength in liquid scintillator $l_{\rm{LS}} = 2\sqrt{R^2_{\rm{LS}}-D^2}$, also the length in water is defined as $l_{\rm{buffer}} = l_{\rm{CD}}-l_{\rm{LS}}$. Figure \ref{fig:trackLengths} shows the construction and the influence of $D$ on the track lengths in LS and water. With JUNO being a spherical detector, a track's orientation in $\theta$ and $\phi$ has a smaller impact on its characteristics than the parameter $D$.
Tracking in liquid scintillator is a challenge, because the light emission is isotropic and not directional. For an extensive track it can be used that all isotropic light emissions along the track add up. Since the travel speed of photons is reduced by the refractive index of the LS $v_g = c_0 / n_{\rm{LS}}$, a forward-moving light front in shape of a cone is formed. As shown in figure \ref{fig:construction}, the construction is similar to a Cherenkov-light cone and carries information about the track's position and direction. The muon track is the central axis of the cone and encloses the opening angle $\theta_\alpha$ with the light front, which is the mantle of the cone. The opposite angle $\theta_c$ is the angle under which the photons were emitted that add up to the fastest light front. The scintillator in JUNO is expected to have a refractive index $n=1.485$. According to figure \ref{fig:construction}, the expected opening angle of the cone is then $\theta_\alpha = \arccos(\frac{1}{\beta n_{\rm{LS}}}) = 47.7^\circ$. This assumes a through-going muon with $\beta = 1$.

\section{The cone model}
\label{sec:conemodel}

\subsection{Description of light propagation}
\label{sec:lightpropa}
\begin{figure}
\centering
	\begin{tikzpicture}[scale=1.14, spy using outlines]
	\draw (0,0) circle [radius=1.77]; \node at (1,-1) {\tiny LS sphere}; 	
	\draw (0,0) circle [radius=1.95]; \node at (1.6,-1.7) {\tiny PMT array};	
	\centerarc[blue](0,0)(-7:175:1.95)
	\centerarc[orange](0,0)(175:353:1.95)
	\draw [rounded corners] (-4.35/2.,-2.2) rectangle (4.35/2.,2.4);
	\begin{scope}[shift={(-0.844278,1.555666)}]
		\tstar{0.025}{0.05}{7}{10}{thick,fill=blue}
	\end{scope}
	\draw [dotted] (-0.844278,1.555666) -- (2.095464,-0.343787); 
	\draw [dotted] (-0.844278,1.555666) -- (-2.081305,-0.015882); 
	\centerarc[black](-0.654422,0.416532)(57:100:0.5)
	\node [scale=0.5] at (-0.575,0.75) {$\theta_\alpha$};
	\draw [thick, red] (-0.654422,0.416532) -- (-0.184986,1.143068); 
	\draw [thick, red] (-0.654422,0.416532) -- (-1.334117,0.951546); 
	\centerarc[red,thick](-0.844278,1.555666)(-32.867678:231.792322:0.776)
	\draw [fill] (-0.184986,1.143068) circle [radius=0.025];
	\draw [fill] (-1.334117,0.951546) circle [radius=0.025];
	\draw [dashed, ->] (-0.5*1.95, 1.2*1.95) -- (-0.335601*1.95,0.213606*1.95); 
	\node at (-0.60,2.055666) {\tiny $\mu$-track};

	\begin{scope}[shift={(4.5,0)}]
    		\draw (0,0) circle [radius=1.77]; \node at (1,-1) {\tiny LS sphere}; 	
		\draw (0,0) circle [radius=1.95]; \node at (1.6,-1.7) {\tiny PMT array};	
		\centerarc[blue](0,0)(-7:175:1.95)
		\centerarc[orange](0,0)(175:353:1.95)
		\draw [rounded corners] (-4.35/2.,-2.2) rectangle (4.35/2.,2.4);
		\draw [dashed, ->] (-0.5*1.95, 1.2*1.95) -- (-0.481803,-0.619182); 
		\node at (-0.60,2.055666) {\tiny $\mu$-track};
		\begin{scope}[shift={(-0.844278,1.555666)}]
			\tstar{0.025}{0.05}{7}{10}{thick,fill=blue}
		\end{scope}
		\draw [dotted] (-0.844278,1.555666) -- (2.095464,-0.343787); 
		\draw [dotted] (-0.844278,1.555666) -- (-2.081305,-0.015882); 
		\centerarc[black](-0.481803,-0.619182)(57:100:0.5)
		\node [scale=0.5] at (-0.402383,-0.285714) {$\theta_\alpha$};
		\draw [thick, red] (-0.481803,-0.619182) -- (0.402799,0.749898); 
		\draw [thick, red] (-0.481803,-0.619182) -- (-1.762615,0.388996); 
		\centerarc[red,thick](-0.844278,1.555666)(-32.867678:34:1.4864)
		\centerarc[red,thick](-0.844278,1.555666)(145.5:153:1.4864)
		\centerarc[red,thick](-0.844278,1.555666)(207.5:231.792322:1.4864)
		\draw [fill] (0.402799,0.749898) circle [radius=0.025];
		\draw [fill] (-1.762615,0.388996) circle [radius=0.025];
  	\end{scope}	
	
	\begin{scope}[shift={(9,0)}]
    		\draw (0,0) circle [radius=1.77]; \node at (1,-1) {\tiny LS sphere}; 	
		\draw (0,0) circle [radius=1.95]; \node at (1.6,-1.7) {\tiny PMT array};	
		\centerarc[blue](0,0)(-7:175:1.95)
		\centerarc[orange](0,0)(175:353:1.95)
		\draw [rounded corners] (-4.35/2.,-2.2) rectangle (4.35/2.,2.4);	
		\draw [dashed, ->] (-0.5*1.95, 1.2*1.95) -- (-0.235205,-2.098773); 
		\node at (-0.60,2.055666) {\tiny $\mu$-track};
		\begin{scope}[shift={(-0.844278,1.555666)}]
			\tstar{0.025}{0.05}{7}{10}{thick,fill=blue}
		\end{scope}
		\draw [dotted] (-0.844278,1.555666) -- (2.095464,-0.343787); 
		\draw [dotted] (-0.844278,1.555666) -- (-2.081305,-0.015882); 
		\centerarc[black](-0.235205,-2.098773)(57:100:0.75)
		\node [scale=0.5] at (-0.125785, -1.565305) {$\theta_\alpha$};
		\draw [thick, red] (-0.235205,-2.098773) -- (1.251795,0.202625); 
		\draw [thick, blue] (-0.235205,-2.098773) -- (-0.099530,-1.888791); 
		\draw [thick, red] (-0.235205,-2.098773) -- (-2.176066,-0.571044); 
		\draw [thick, blue] (-0.235205,-2.098773) -- (-0.431648,-1.944145); 
		\centerarc[red,thick](-0.844278,1.555666)(-32.867678:19.8:2.493)
		\draw [fill] (-0.294101,-1.745395) circle [radius=0.025];
		\draw [dotted] (-0.294101,-1.745395) -- (0.419836,-2.206691); 
		\draw [dotted] (-0.294101,-1.745395) -- (-0.634283,-2.177571); 
		\centerarc[red,thick](-0.294101,-1.745395)(231.792322:327.132322:0.245)
		\draw [fill] (1.251795,0.202625) circle [radius=0.025];
  	\end{scope}
	\spy [purple, thick, draw,height=2.5cm,width=2.5cm,magnification=2] on (9.9,-1.87) in node [fill=white] at (10, 1.2);
		
	\end{tikzpicture}
	\caption{\label{fig:coneSphereInDetector}The evolution of the first light front for a muon traveling through the central detector of JUNO. The muon track is represented by a dashed line and the entry point into the LS with a black star. The left plot shows the light front after about $1/3$ of the track length. The dotted lines separate the PMT array into an area that was or will be hit by the sphere-part and the one that will see light from the cone-part of the model. The transitions between the two categories of light fronts are marked with black dots. The middle plot depicts the evolution of the light front after the muon travelled about $2/3$ of its track. On the right, the muon already left the CD, but some PMTs are yet to be struck by the light front. In addition, the spherical scintillation around the exit point is marked. At this time, the cone's apex consists of Cherenkov light from traversing the waterbuffer. The insert is a zoomed-in view on the exit point. The circle segment models the fastest scintillation photons from the exit point out of the LS.}
\end{figure}
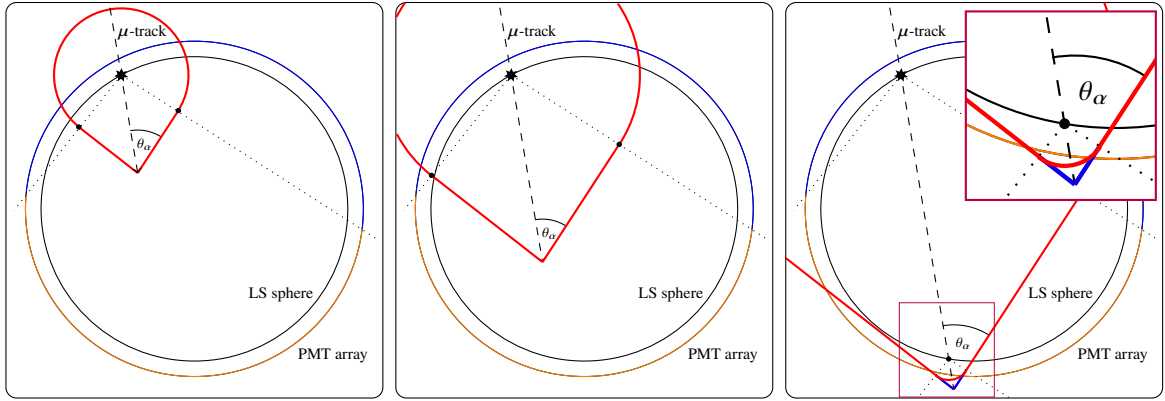

The focus for the light propagation model is on the scintillation light, which is dominant for all events that hit the CD. According to figures \ref{fig:construction} and \ref{fig:coneSphereInDetector}, the first photons traveling behind the muon can be described with a conical shape. The model has to be extended, when it is applied to a real detector. The track starts when the muon enters the LS and builds the cone in the forward direction. Due to that, the PMTs behind the muon around the entry point cannot receive light from the cone. They collect photons from the isotropic emission along the very first bit of track inside the LS.
In order to model the signal, the cone model is extended with a backward sphere. The same effect is present on the apex of the cone, when the muon exits the acrylic sphere. After leaving the LS, no more scintillation light is produced by the muon and the apex of the cone is smoothed out by a sphere around the last bit of tracklength in the LS. In contrast to the entry point, the Cherenkov light, created after the muon leaves the LS, is directed towards the PMTs. This allows to continue the conical model for the PMTs close to the muon's exit point, but with an opening angle according to the refraction in water instead of that in LS. For some track orientations with large D, it is also possible that PMTs around the entry point see light directly from the Cherenkov light produced in the waterbuffer before the muon enters the liquid scintillator. Thus, their hit time is much earlier than predicted with the scintillation model. Instead of including this effect in the model, the PMTs are removed from the fit as described in section \ref{sec:testprocedure}. Cherenkov light is also produced in the LS together with the scintillation light. A separation of Cherenkov- and scintillation-photons from inside the LS is not needed with the cone model.
Most PMTs receive their first hit from scintillation photons and in those cases, where Cherenkov hits are earlier, they are only separated from scintillation hits by less than \unit[3]{ns}.
Section \ref{sec:testprocedure} will show in detail that this is on the same order of magnitude as the reconstructed hit times from the expected waveforms.

\subsection{Implementation of the model} 
\label{sec:implementation}
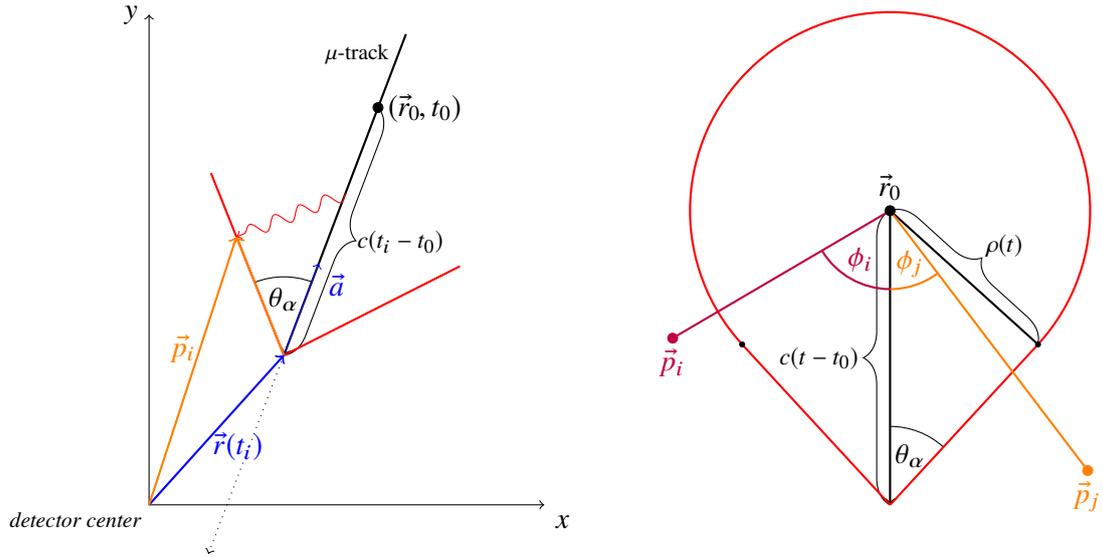
\begin{figure}
\centering
	\begin{tikzpicture}[scale=1.3]
	\draw [<->] (0,5) node [left] {$y$} -- (0,0) node [below left, scale=0.75] {\emph{detector center}} -- (4,0) node [below right] {$x$}; 
	\begin{scope}[shift={(-1.2,0)}]
		\draw [decorate,decoration={brace,amplitude=10pt},yshift=0pt]
		(3.517554,4.051519) -- (2.564300,1.525394) node [black,midway, right, yshift=-4pt, xshift=6pt] 
		{\footnotesize $c(t_i-t_0)$};
		\draw [dotted,->] (3.8, 4.8) -- (1.8, -0.5); 
		\draw [thick, ->] (3.8, 4.8) -- (2.564300,1.525394); 
		\draw [thick, red] (2.564300,1.525394) -- (1.826247,3.384232); 
		\draw [thick, red] (2.564300,1.525394) -- (4.346383,2.433237); 
		\draw [fill] (3.517554,4.051519) circle [radius=0.05]; \node [right] at (3.517554,4.051519) {$(\vec{r}_0, t_0)$}; 
		\node [scale=0.75] at (3.3,4.6) {$\mu$-track};
		\draw [->, blue] (2.564300,1.525394) -- (2.917357,2.460996) node [below right] {$\vec{a}$}; 
		\centerarc[black](2.564300,1.525394)(69.325575:111.655575:0.8)
		\node at (2.56, 2.1) {$\theta_\alpha$};
		\coordinate (apex) at (2.564300,1.525394);
		\coordinate (pp) at (2.084566,2.733639);
		\draw [orange, thick, ->] (apex) -- (pp);
		\draw [red,->,decorate,decoration=snake] (3.199868,3.176470) -- (pp);
	\end{scope}
	\draw [blue, thick, ->] (0,0) -- (apex);
	\draw [orange, thick, ->] (0,0) -- (pp);
	\node [blue] at (0.9, 0.6) {$\vec{r}(t_i)$};
	\node [orange] at (0.35, 1.6) {$\vec{p_i}$};
	
	\begin{scope}[shift={(7.5,3.0)}, rotate=-90]
		\coordinate (center) at (0,0); 
		\coordinate (apexx) at (3,0); 
		\coordinate (transup) at (1.363634,1.497039); 
		\coordinate (transdown) at (1.363634,-1.497039); 
		\coordinate (pmta) at (2.65, 2); 
		\coordinate (pmtb) at (1.3, -2.2); 
		\draw [thick, ->] (center) -- (apexx);
		\draw [decorate,decoration={brace,amplitude=10pt, mirror},yshift=0pt]
		(center) -- (apexx) node [black,midway, left, yshift=0pt, xshift=-8pt] 
		{\footnotesize $c(t-t_0)$};
		\draw [decorate,decoration={brace,amplitude=10pt},yshift=0pt]
		(center) -- (transup) node [black,midway, above right, xshift=4pt, yshift=4pt] 
		{\footnotesize $\rho(t)$};
		\centerarc[black](apexx)(90+47.67:180:0.8)
		\node at (2.5, 0.2) {$\theta_\alpha$};
		\draw [thick, red] (apexx) -- (transup); 
		\draw [thick, red] (apexx) -- (transdown); 
		\centerarc[thick, red](center)(47.67:360-47.67:2.025)
		\draw [thick] (center) -- (transup);	
		\draw [fill] (transup) circle [radius=0.025];
		\draw [fill] (transdown) circle [radius=0.025];
		\draw [fill, orange] (pmta) node [below] {$\vec{p}_j$} circle [radius=0.05];
		\draw [fill, purple] (pmtb) node [below] {$\vec{p}_i$} circle [radius=0.05];
		\draw [thick, orange] (center) -- (pmta);
		\draw [thick, purple] (center) -- (pmtb);
		\centerarc[orange, thick](center)(0:36.5:0.8)
		\centerarc[purple, thick](center)(-59:0:0.8)
		\node [orange] at (0.54, 0.2) {$\phi_j$};
		\node [purple] at (0.5, -0.3) {$\phi_i$};
		
		\draw [fill] (0,0) node [above] {$\vec{r}_0$} circle [radius=0.05]; 
	\end{scope}
	
	\end{tikzpicture}
	\caption{\label{fig:coneSphereConstruction}\emph{Left:} Implementation of the cone model with a vector that points to the apex of the cone $\vec{r}(t)$ and an inverse muon-track direction unit vector $\vec{a}$. The detector center is at $(0,0)$ and $\vec{r}_0$ corresponds to $t_0$, giving the entry point into the LS and its time. A PMT at position $\vec{p_i}$ would be struck by the photons depicted that were emitted at a certain point along the track. According to equation \ref{eq:theta_alpha}, the corresponding cone would have an opening angle $\theta_\alpha$ between the muon direction and light front. The plot shows a snapshot at the time $t_i$ when the PMT at position $\vec{p_i}$ first detected light and the muon was at position $\vec{r}(t_i)$. In this fashion, a cone with a certain opening angle $\theta_\alpha$ can be constructed from the muon track hypothesis for every PMT. \emph{Right:} Implementation of the sphere around the entry point to close the bottom of the cone. It can be derived from the opening angle $\theta_\alpha$ as well. A PMT, struck at time $t_i$ by the spherical light front around the entry point into the LS $(\vec{r}_0, t_0)$, defines the radius $\rho(t_i)$. With the track length $c(t_i-t_0)$ to the apex of the continuously merging cone the opening angle can be calculated. For two arbitrary PMT positions $\vec{p}_i$ and $\vec{p}_j$ the angles $\phi_i$ and $\phi_j$, respectively, for category-weighting are also shown.}
\end{figure}

In contrast to other fastest-light-based muon track reconstructions \cite{Abe:2014yda, borexMuon:2011}, this approach implements the geometrical models of the light cone and sphere. Thus, the figure of merit in the fit is not a time residual, but the opening angle $\theta_\alpha$ of the first light cone. Incorporating the physical dependencies of light propagation, also the sphere that models backward moving light can be defined distinctly by this angle, as constructed in figure \ref{fig:coneSphereConstruction}. In the following, four signal categories are introduced that model different aspects of light propagation in the detector:

\paragraph{(1) Forward Cone.}
The track parameters are the entry point into the LS, the time of entry, and the muon's direction $\vec{d}$. This information is transformed into a description of the time dependent position of the muon in the detector. According to figure \ref{fig:coneSphereConstruction}, this also coincides with the apex of the cone
\begin{equation}
\vec{r}(t) = \vec{r}_0 - c(t-t_0)\vec{a},
\end{equation}
with $\vec{r}_0$ being the entry point into the LS, $t_0$ the entry time, and $ \vec{a} = -\vec{d}$ a unit vector of the inverted direction of the muon. This allows to track the muon through the detector and to build the model of the first photons forming a conical surface. Every point $\vec{x}$ on the surface of a cone around the track with opening angle $\theta_\alpha$ can be described by
\begin{equation}
\frac{\vec{x}-\vec{r}(t)}{|\vec{x}-\vec{r}(t)|}\cdot \vec{a} = \cos(\theta_\alpha).
\end{equation}
The available data are the static PMT positions $\textbf{P} = \{\vec{p_1},\vec{p_2},...,\vec{p_n}\}$ and their set of first hit times $\textbf{T} = \{t_1, t_2,..., t_n\}$ for each event. Under the assumption that the PMT was hit by the earliest photons possible, the first light cone can be constructed with an opening angle of
\begin{equation}
\label{eq:theta_alpha}
\theta_{\alpha,i} = \arccos\left( \frac{\vec{p}_i-\vec{r}(t_i)}{|\vec{p}_i-\vec{r}(t_i)|}\cdot \vec{a} \right).
\end{equation}
Thus, the opening angle is defined as the angle between the track direction and the connection between the position of the muon $\vec{r}(t_i)$ at hit time $t_i$ and the PMT at position $\vec{p}_i$.
Since both vectors of the product are normalized, this function will always yield an angle between 0 and $\unit[180]{^\circ}$, even for track parameters far off the true values.
Figure \ref{fig:coneSphereConstruction} gives an example of the cone construction. It shows an event at a time $t_i$, when a PMT at position $\vec{p}_i$ is first hit. The track hypothesis $\vec{q}$ contains $\vec{r}_0$, $t_0$, and $\vec{a}$. The position of the muon at the apex of the cone $\vec{r}(t_i)$ can be calculated with the track hypothesis. The cone according to the hit times data $t_i$ and the track $\vec{q}$ is then calculated. Its opening angle $\theta_\alpha$ is enclosed by the link vector between $\vec{r}(t_i)$ and $\vec{p}_i$ and the inverse direction $\vec{a}$.
\paragraph{(2) Backward Sphere.}
The sphere model to describe light moving behind the muon is described in a similar fashion. It has to merge continuously with the cone and is centered around the entry point $\vec{r}_0$. The radius of this light sphere depends on the opening angle $ \theta_\alpha$ via
\begin{equation}
\rho(t) = c(t-t_0)\cdot \sin(\theta_\alpha).
\end{equation}
If a PMT was hit by this sphere, the radius is specified and the corresponding opening angle is calculated by
\begin{equation}
\theta_{\alpha,i} = \arcsin\left( \frac{|\vec{r}_0 - \vec{p}_i|}{c(t_i-t_0)} \right).
\end{equation}
For certain arguments, the $\arcsin$ can run out of bounds. This happens if the track parameters are too far from the true values or when the scintillation model is not sufficient. In those cases, the $\arcsin$ is extended to return the value of $\pi$ or 0, respectively, to reflect the low probability, while maintaining a smooth function.
This model is very idealized because it assumes that exactly at the entry point into the LS, enough light is produced to reach all PMTs crossed by the sphere-front. According to figure \ref{fig:coneSphereInDetector}, this can be a substantial part of the whole signal. One possibility to take the finite tracklength for production of enough photons into account, is to shift the effective point of emission along the track. The reconstruction takes $ {\vec{\tilde r}_0} = \vec{r}_0 - l\cdot \vec{a}$ with $l=\unit[20]{cm}$ as the effective center of the backwards sphere.

\paragraph{(3) Cherenkov Cone and (4) Forward Sphere.}
In addition, two more categories of signal are introduced in the algorithm. Around the last point of the track in LS, another sphere is built to model the scintillation light on the cone's apex when it enters the waterbuffer again. It is implemented like the backward sphere.
The forth category of signal models the Cherenkov cone in the waterbuffer after the muon exited the LS. Its implementation is identical to the normal cone explained above and it is only applied to those PMTs that fall into category (3).

\begin{figure}[htbp]
\centering
\includegraphics[width=\textwidth]{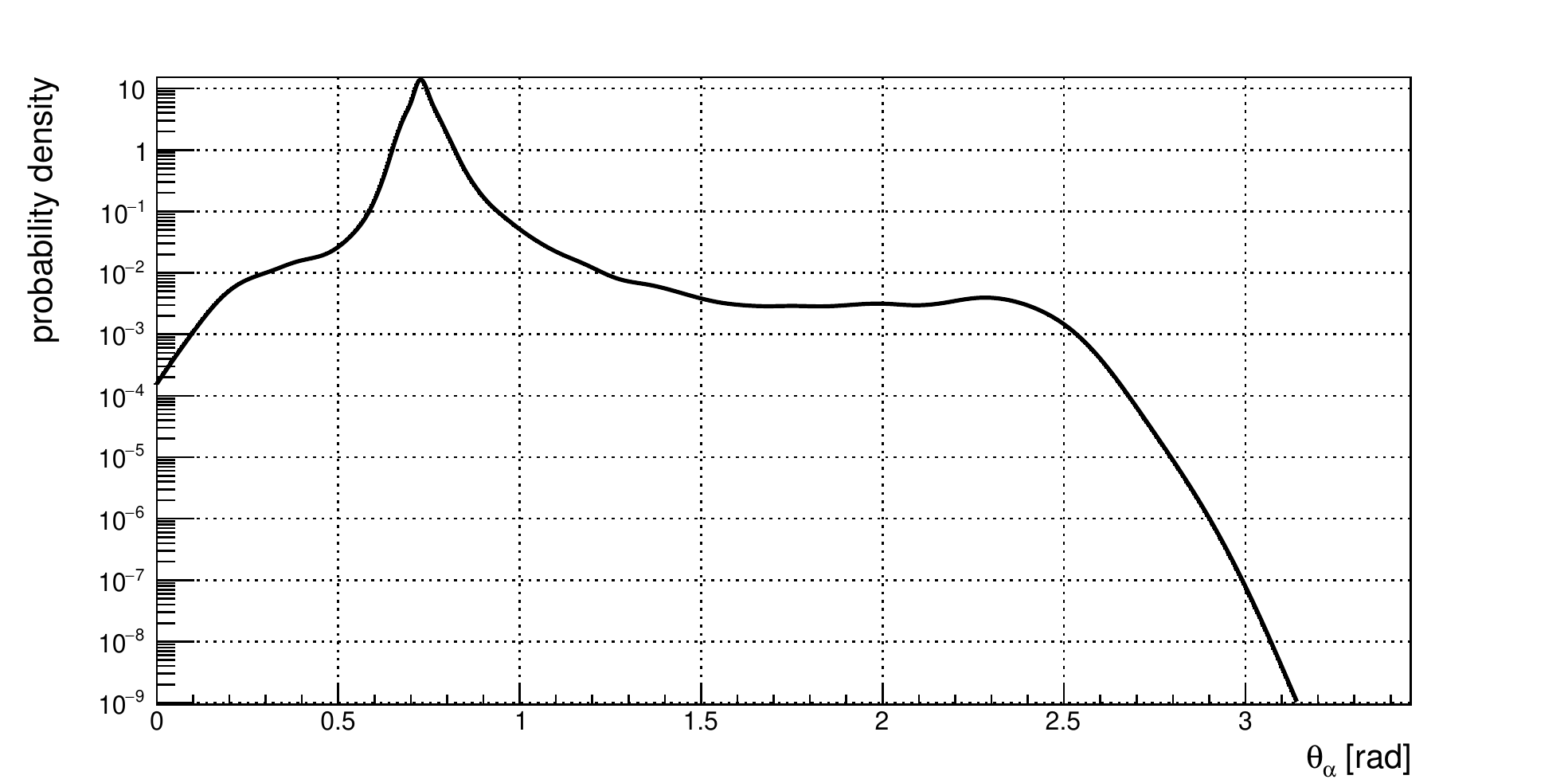}
\caption{\label{fig:pdf} Probability density function of $\theta_\alpha$ for $D=\unit[0]{m}$ for the LPMT cone-part of the model. It is produced by plugging-in the first hit times from the Monte Carlo simulation of muon tracks into the reconstruction model. It was smoothed over the whole domain of definition with adaptive kernel density estimators \citep{scott:1992}.}
\end{figure}

\paragraph{Likelihood function.}
\label{par:llhfunc}
The actual fit is performed with a likelihood function minimized by MINUIT\cite{minuit:1975}. For each function evaluation, all selected PMTs have to be assigned to one of the four categories explained above. For each PMT, the angle between the muon-track direction and the connection between PMT and entry point $\vec{r}_0$ can be used to assign it either to the cone (1) or the backward sphere (2) category. The same calculation is done with the exit point $\vec{r}_{exit}$ to further categorize PMTs from the cone part to case (3) or (4).
In order to achieve a smooth likelihood function, the transitions between categories have to be continuous. In addition to that, the track's distance from center heavily influences the share of PMTs that fall into the different categories. For this reason, the opening angle and its probability for all four categories are calculated and weighted by a transition function. In the algorithm, the weighting function was chosen to be a normalized error function. For example, the weight for the $i$-th PMT for category (2) is given by
\begin{equation}
w_{2,i} = \frac{1}{2} \left( 1+ \erf \left(s \cdot \Delta \phi_i \right) \right),
\end{equation}
where $\Delta\phi_i = \phi_i - \Phi_{\rm LS}$, $\phi_{i}$ being the angle enclosed by the direction of the track and the link between the PMT and the LS entry point $\vec{r}_0$. Two examples for $\phi_{i}$ are shown in the right panel of figure \ref{fig:coneSphereConstruction}.
$\Phi_{\rm LS} = \arccos\left(\frac{1}{n_{\rm LS}} \right)$ is the constructed transition angle in LS and $s$ is a constant factor to scale the transition width. The scaling factor $s$ has only a negligible effect on the reconstruction and was chosen to be $s=25$ for all weights. If $\phi_i = \Phi_{\rm LS}$, the PMT is exactly at the transition between the cone-part and the sphere-part and will receive a weight of $0.5$ for both  categories. For $\phi_i > \Phi_{\rm LS}$, the weight $w_{2,i}$ for the sphere-part increases up to $1$.
In the same fashion, the weight for category (3) is defined and the weight for (1) is given by $ w_1 = 1-w_2-w_3 $. Category (4) is a subclass that can only be applied to the selection of PMTs in (3). The transition is also modeled by an error function and increases the weight of the Cherenkov model when the PMT hit time is significantly earlier than predicted by the scintillation model (3).
The complete log-likelihood function sums over all $n_{\rm PMT}$ PMTs that detected light and were selected for the fit. For each PMT a cone and a sphere is calculated according to its data $(\vec{p}, t)$ and the track hypothesis $\vec{q}$ as explained before. The complete function is given by
\begin{equation}
-2\log\mathcal{L} = -2 \sum^{n_{\rm PMT}}_{i=0}\ln f_{i}(\theta_{\alpha,i};\vec{q}),
\end{equation}
with the probability function
\begin{equation}
\begin{split}
f_{i}(\theta_{\alpha,i};\vec{q}) & ={} w_{2,i}(\vec{q}) P_2(\theta_{\alpha,i}|\vec{q}) \\
 & + w_{3,i}(\vec{q}) \left[ w_{4,i} P_4(\theta_{\alpha,i}|\vec{q}) + (1-w_{4,i}(\vec{q})) P_3(\theta_{\alpha,i}|\vec{q})\right] \\
 & + (1-w_{2,i}-w_{3,i}) P_1(\theta_{\alpha,i}|\vec{q}).
\end{split}
\end{equation}
The probabilities $P_j$ are evaluated from the pre-calculated, normalized angle distributions. For each of the four categories, there are 18 probability functions for each \unit[1]{m} step in the distance from center $D$. For a given set of track parameters $\vec{q}$, the probability is linearly interpolated between the values of the functions of the two enclosing values for $D$. An example of a probability density function for the cone-part can be seen in figure \ref{fig:pdf}. To account for the difference in light collection and geometrical placement there are two dedicated sets of PDFs for the LPMT and the SPMT systems. The performance differences within the LPMT array between dynode- and MCP-PMTs are sufficiently small to justify using the average PDF for the full LPMT system.
The functions were obtained by plugging the Monte Carlo truth track parameters of simulated events into the model and extracting the distribution of $\theta_\alpha$.
For this purpose a set of 3000 muons was simulated. They are set to penetrate the central detector at distances from center $D=\unit[0]{m}$ to $\unit[17]{m}$ in \unit[1]{m}-steps, with inclinations between $\theta = \unit[180]{^\circ}$ (straight down) to $\unit[100]{^\circ}$ and at 10 orientations in spherical $\phi$.
The true hit times were smeared according to the PMT time resolution in order to include those effects in the probability distributions. Incorporating adaptive kernel density estimators \cite{scott:1992}, the distributions were smoothed over the full range of possible angles to increase the stability of the fit.
The fit has 5 parameters in total. Two define the track entry point on the LS sphere, one the corresponding entry time $t_0$, and two more the muon track's direction unit vector.
The starting parameters for the fit are provided by a fast and simple tracking algorithm. In the first step, the distance from center $D$ is extracted from the event's first hit time distribution. The time from the first PMT being hit until the last PMT being hit is linearly correlated with the track's distance from center. According to section \ref{sec:char}, we can directly get the track length in the CD from $D$ via $l_{CD} = \sqrt{R_{CD}^2 - D^2}$. The starting point of the track is the charge weighted sum of positions of fired PMTs within \unit[2.5]{ns} of the first fired PMT. The starting point and track length confine a window in space and time on the PMT array for the exit point. The window is gradually increased until a group of 6 PMTs is inside, in order to charge-weight their positions to obtain an exit point. The entry- and exit-points define a straight track from which the seeding parameters for the fit are calculated.

\section{Test procedure}
\label{sec:testprocedure}
The reconstruction algorithm has to be tested with simulated data, because JUNO is currently under construction and data taking has not yet started. This allows to study the model for a perfect detector and the impact of different stages of the detector response.
\paragraph{Detector simulation.}
In the first step, a full Geant4 \cite{geant4:2002} simulation of the muon events is performed. The detector geometry includes a fully detailed model of the central detector submerged in the waterpool. The forseen PMT placement with all holding structures is modeled, as well as the \unit[12]{cm} thick acrylic sphere that holds the liquid scintillator. The optical model is build according to the proposed detector design \cite{cdr:2015}. In addition, also the quantum efficiency and collection efficiency of the PMTs is modeled. The collection of registered photoelectrons of an average muon event contains about $\unit[10^6]{entries}$.

\begin{figure}[htbp]
\centering
\includegraphics[width=\textwidth]{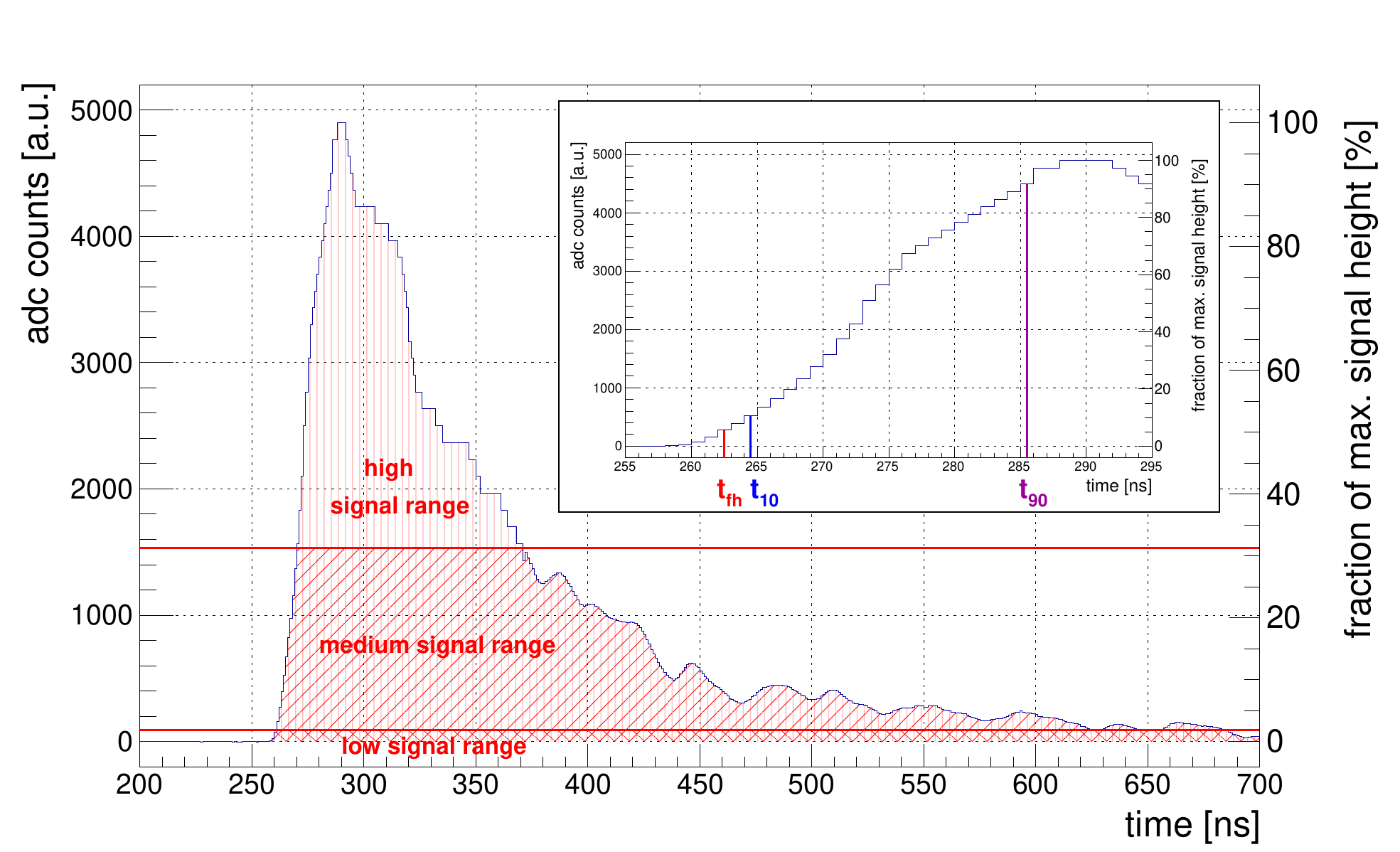}
\caption{\label{fig:waveform} Simulated waveform of one LPMT for a muon event. The first hit time $t_{\rm fh}$ is defined as the time when the rising edge crosses the threshold of 6\% of the total signal height. Additionally, the $t_{10}$ and $t_{90}$ points are marked, which are used to define the signal's rise time.
The different ranges of the 3x8bit FADC are visible through the change of resolution for the high, medium and low signal regions.
}
\end{figure}

\paragraph{Electronics simulation.}
The second step of simulation models the LPMT response and their FADC electronics. The LPMT simulation adds an estimated darknoise of \unit[20]{kHz} to the signal. In a readout window of \unit[1250]{ns} with a \unit[1]{GHz} sampling, a pulse is added for every simulated hit according to its arrival time and charge. According to the different kinds of PMT explained in section \ref{sec:intro}, a TTS is assigned to every LPMT. Every hit time is smeared according to the LPMT's TTS and the charge is varied according to the LPMT resolution. For each hit, a pulse is created with the shape of a single p.e. log-normal function scaled by the hit's charge. The parameters for this function are extracted from PMT measurements at Daya Bay \cite{Jetter:2012} and dedicated measurements on 20-inch PMTs. Afterwards, all pulses on one LPMT are added to build its raw waveform. Finally, the simulation samples the raw waveform with a 3x8bit FADC that can provide a dynamic range of 1600 photoelectrons. It is built in a way, that it can keep approximately the same relative resolution over the whole dynamic range. In figure \ref{fig:waveform} the simulated effect can be seen by the coarser ADC trace in the high signal range and the much smoother curve in the mid and lower range.
The output is expected to be of the same format as the real data will be.
For the SPMT-array the main factor is the TTS. Since they will not produce waveforms, but deliver hit times and charge directly, their simulation consists of a Gaussian hit-time smearing according to their TTS.
\paragraph{Signal reconstruction}
When working with waveforms, the concept of the first-hit times translates to the starting time of the waveform. In order to reconstruct this time consistently over the large dynamic range of the muon signals, we use the principle of a constant-fraction discriminator, as shown in figure \ref{fig:waveform}.
This approach correlates the first-hit time to the time, when the rising edge of the waveform passes a threshold that is determined as a relative fraction of the signal height. A study of the simulated waveforms showed that a 6\% threshold gives the most stable results. The time is linearly interpolated between the two waveform samples below and above the threshold.
The achieved first hit-time resolution over all \unit[20]{inch} PMTs is \unit[3.5]{ns}. This is better than the average TTS of the LPMT system. While the TTS describes the resolution for a single hit, the rising edge consists of hundreds of hits within a few nanoseconds. Due to the high statistics of hits at the same time, it is possible to extract the first hit time with a better resolution than the PMT's TTS.
Additionally, the rise time of the waveform is extracted to aid the PMT selection in the fit. It is defined by the time difference between the muon waveform exceeding 10\% and 90\% of height of the rising edge. Those times are also extracted by linear interpolation between the two samples around each threshold. The charge reconstruction is performed by integration of the waveform over the entire readout time after baseline correction.

\paragraph{Signal cleaning.}
\begin{figure}[htbp]
\centering
\includegraphics[width=0.8\textwidth]{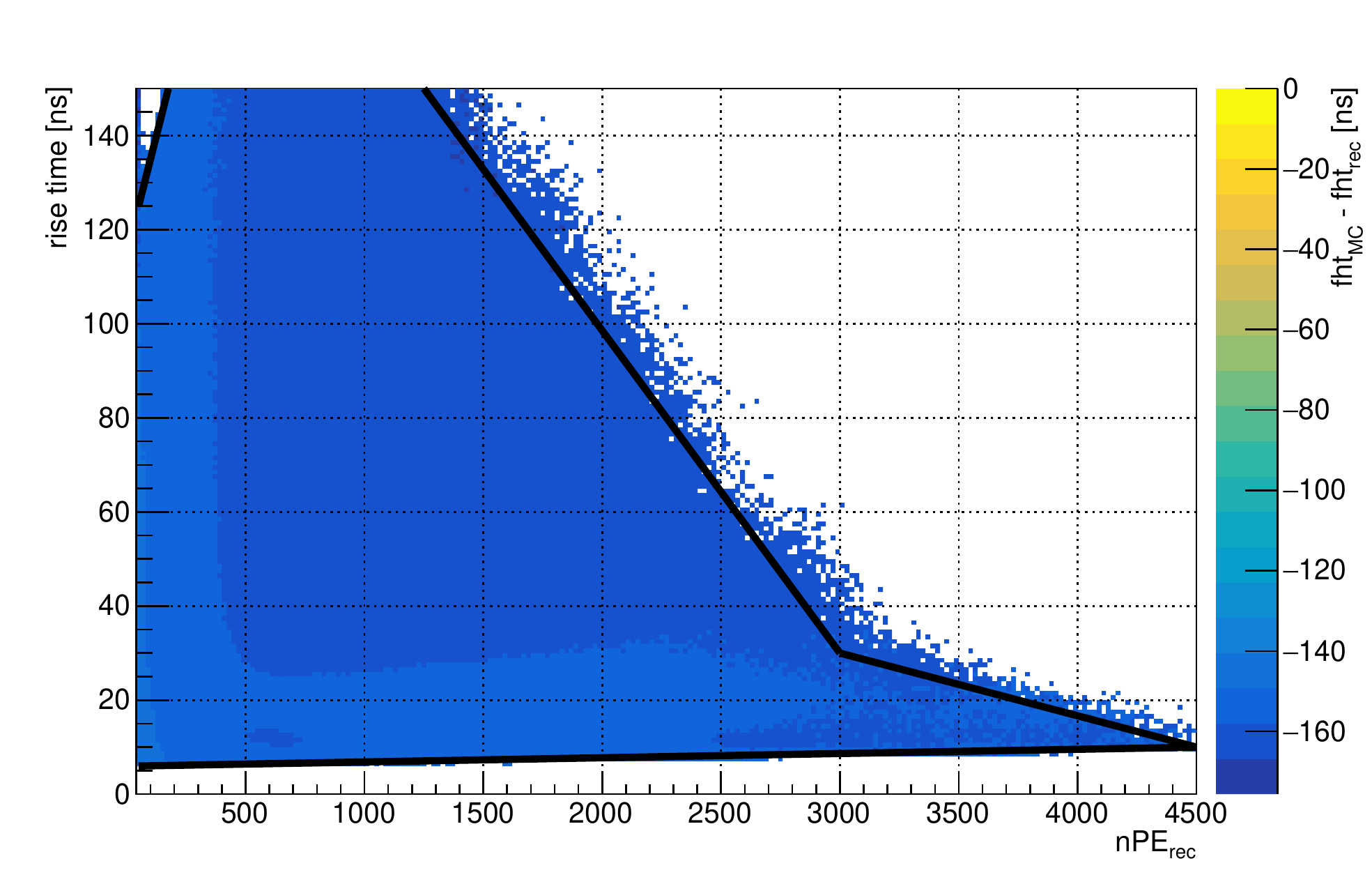}
\caption{\label{fig:2drtcut} The difference between simulated and reconstructed first hit time, in dependence of the reconstructed number of photoelectrons and the rise time. Only the region is shown where the analysis is performed. The black lines indicate a two-dimensional cut to further remove PMTs with a less reliable reconstructed first hit time. The PMTs within the black lines all feature a very similar shift in first hit time, which reflects the good waveform reconstruction performance.
}
\end{figure}
The aforementioned high PMT coverage of (75 + 2.5)\% is due to the dense PMT array. The mean distance between neighbouring LPMTs is less then \unit[60]{cm} center-to-center and it is even lower, when taking both LPMT and the interlaced SPMT into account.
The very high number of channels allows for a strict removal of PMTs with a high uncertainty on the reconstructed first hit time, while maintaining sufficient information for precise tracking. The information of rise time and charge is used to remove some PMTs from the selection the fit will run on. The risetime and charge are arranged in a two dimensional table that was created with the truth information from a sample of through-going muons. Cherenkov light produced in the LS will travel slightly in front of the scintillation light front but it is considerably less intense. This effect can lengthen the rise time and by that also worsen the accuracy of the first hit time reconstruction. Due to that, PMTs with a rise time longer than \unit[150]{ns} are removed from the selection directly. Low charge PMTs have an increased probability to have a first-hit time from reflected or otherwise indirect light. Those photons do not fit the model described above. A general low-charge cut removes PMTs with a charge of less than \unit[50]{p.e.} After those two loose cuts, the two dimensional table, shown in figure \ref{fig:2drtcut}, is used to further remove PMTs. High-charge PMTs have to feature a faster rise time, while lower-charge PMTs have a wider window of allowed rise times.
The high density of PMTs also gives a handle on outlier PMTs that show a reconstructed first hit time due to dark noise or reflected light. According to the cone model, the surface of first light intersects the PMT array continuously. If a PMT was fired by this light front, its hit time cannot deviate much from its direct neighbour's hit times. For each PMT, the mean hit time of its six neighbours is calculated and it is removed, if the difference to its own hit time is more than \unit[5]{ns}.

\paragraph{Test sample.}
In order to get a clear picture of the algorithms characteristics, we test it with a synthetic sample of 5900 simulated muon events. They were simulated according to the procedure explained in section \ref{par:llhfunc}. In this fashion, it is possible to identify areas of the detector where the reconstruction performs worse and to improve it accordingly.
The mean energy of muon events in the central detector is expected to be \unit[215]{GeV} \cite{yb:2015}. In this energy range they will traverse the whole detector and exit again at the bottom. Thus, the muons in the test sample were simulated with this energy. 
In order to test the impact of the detector response, the algorithm is evaluated both with and without electronics simulation.

\section{Tracking performance}
\label{sec:performance}
As explained in section \ref{sec:char}, the track's distance from center $D$ has the main influence on the event characteristics. For this reason, the performance of the reconstruction algorithm is presented against the track's true distance from the center. The benchmark quantities are the deviation in distance from center $\Delta D = D_{\rm sim} - D_{\rm rec}$ and the angle $\alpha$ between the true and the reconstructed track.

\begin{figure}[htbp]
\centering
\subcaptionbox*{LPMT $\Delta D$}{\includegraphics[width=0.40\textwidth]{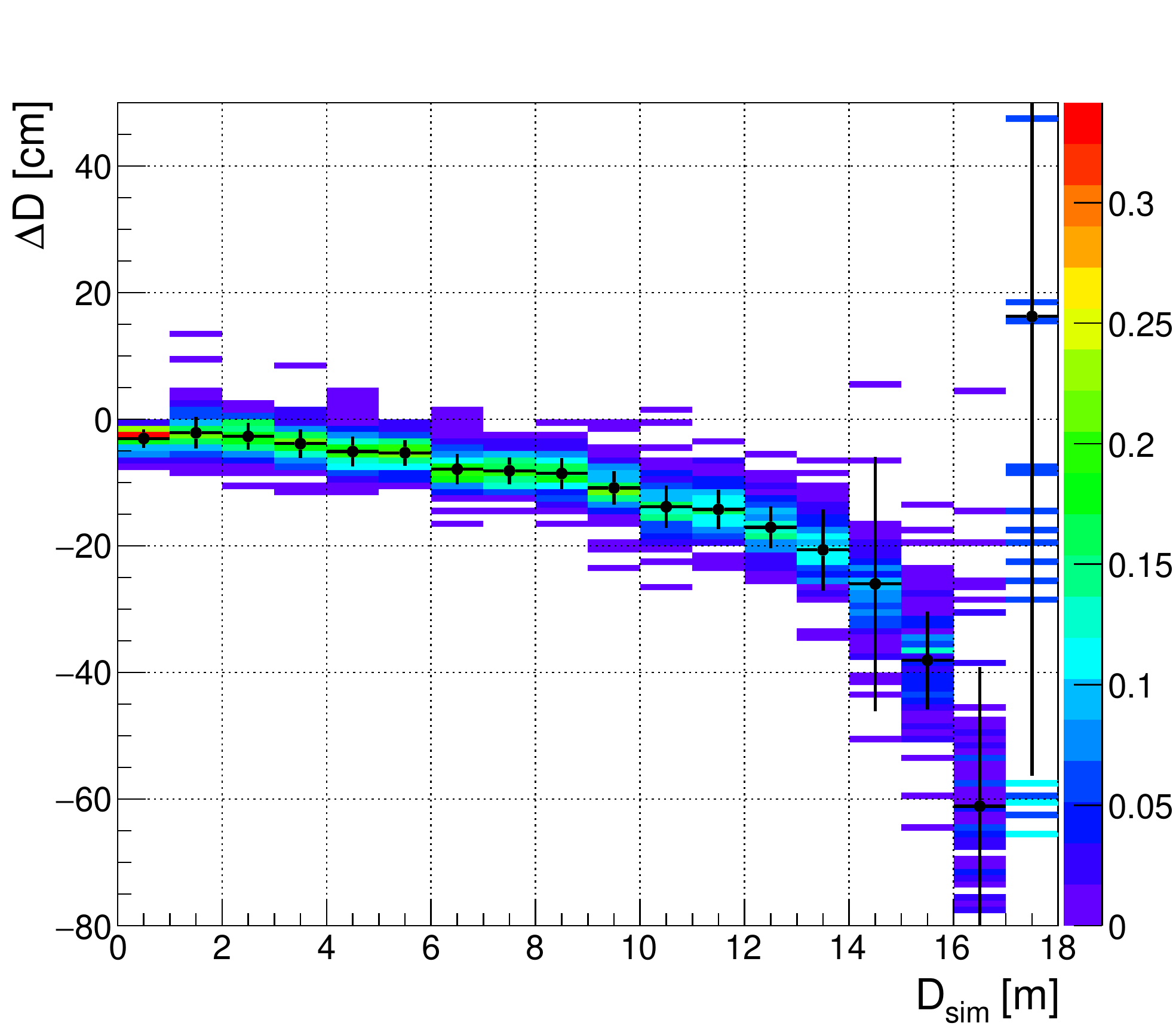}}%
\subcaptionbox*{LPMT $\alpha$}{\includegraphics[width=0.40\textwidth]{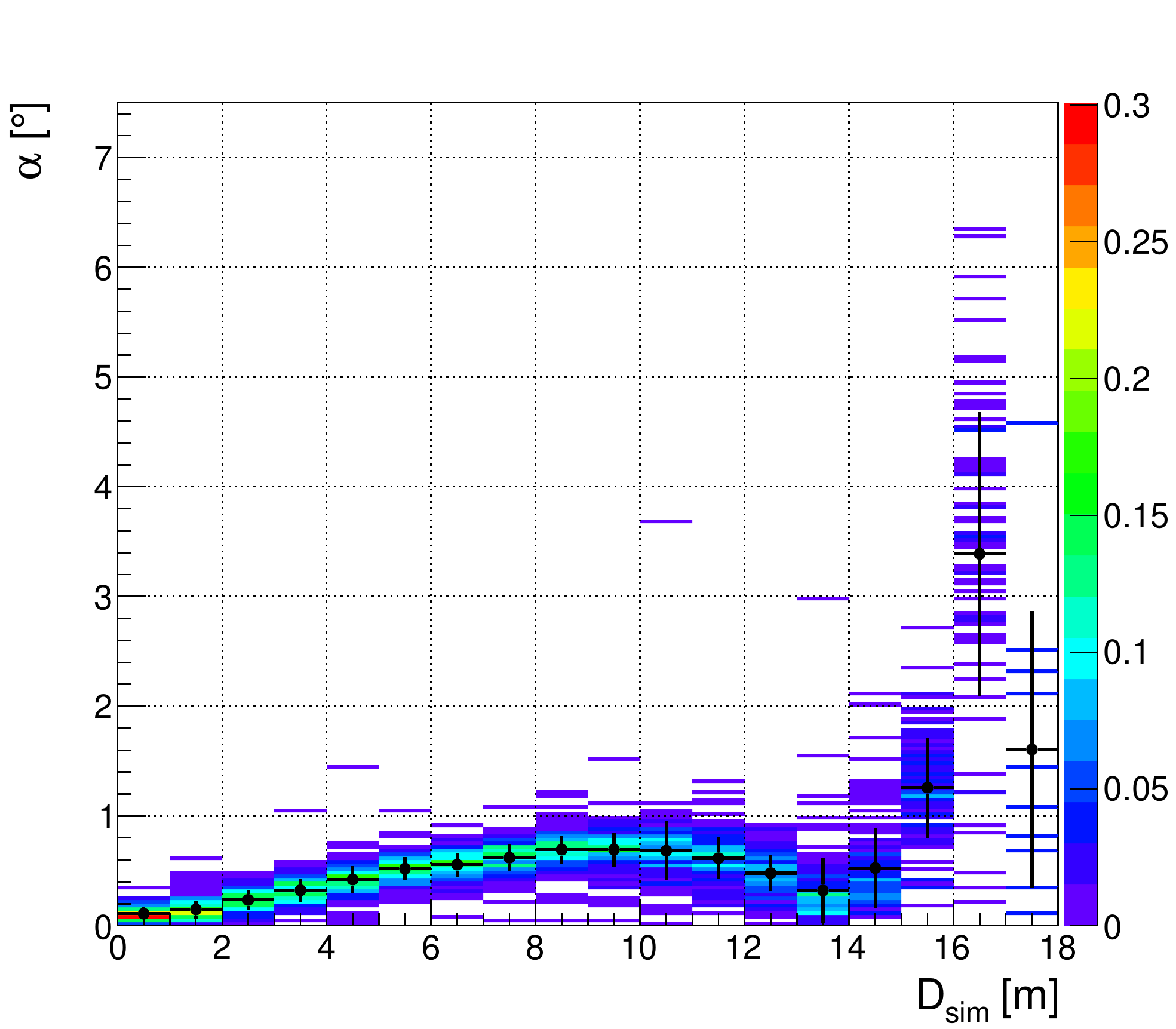}}%
\hfill
\subcaptionbox*{SPMT $\Delta D$}{\includegraphics[width=0.40\textwidth]{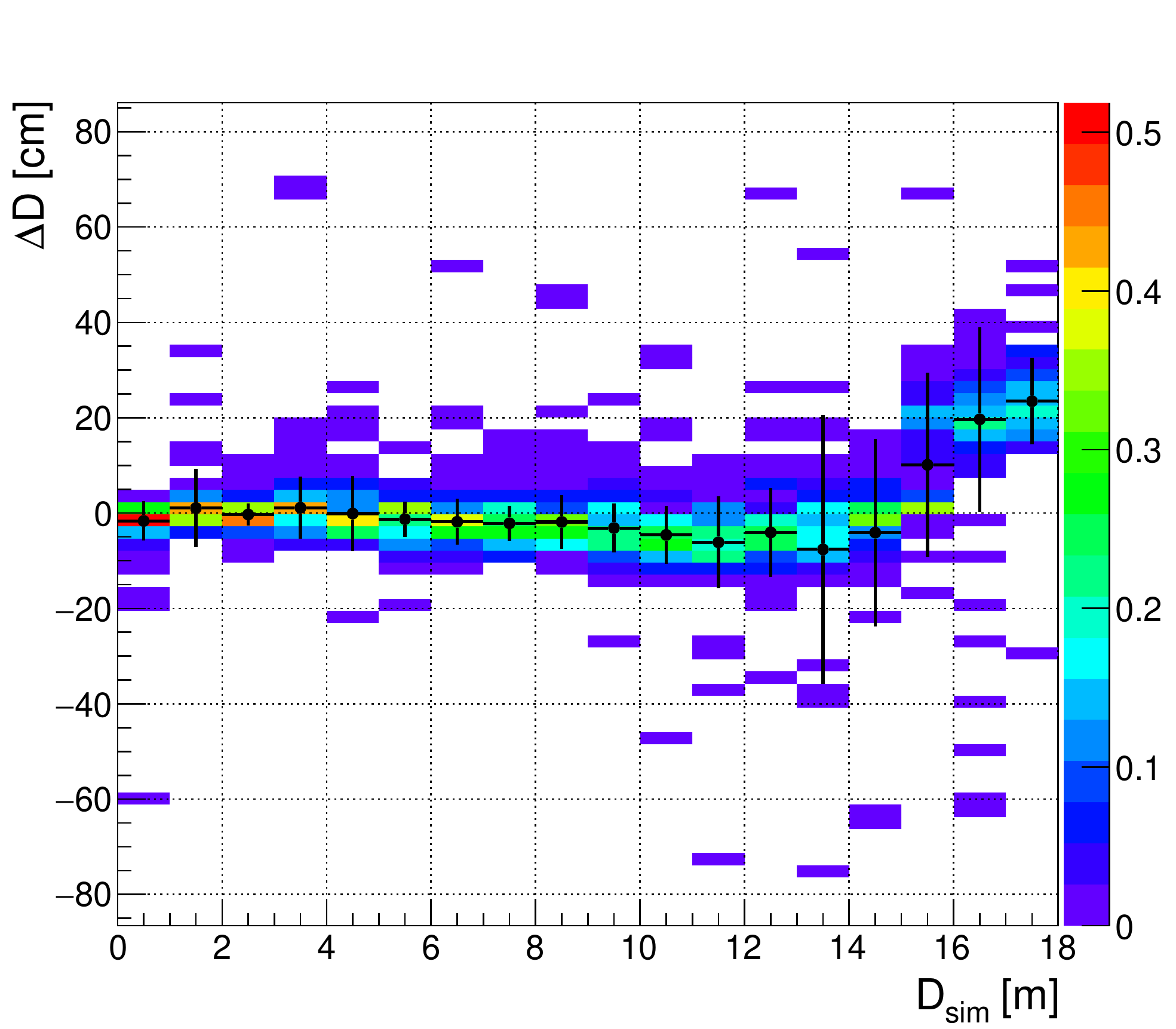}}%
\subcaptionbox*{SPMT $\alpha$}{\includegraphics[width=0.40\textwidth]{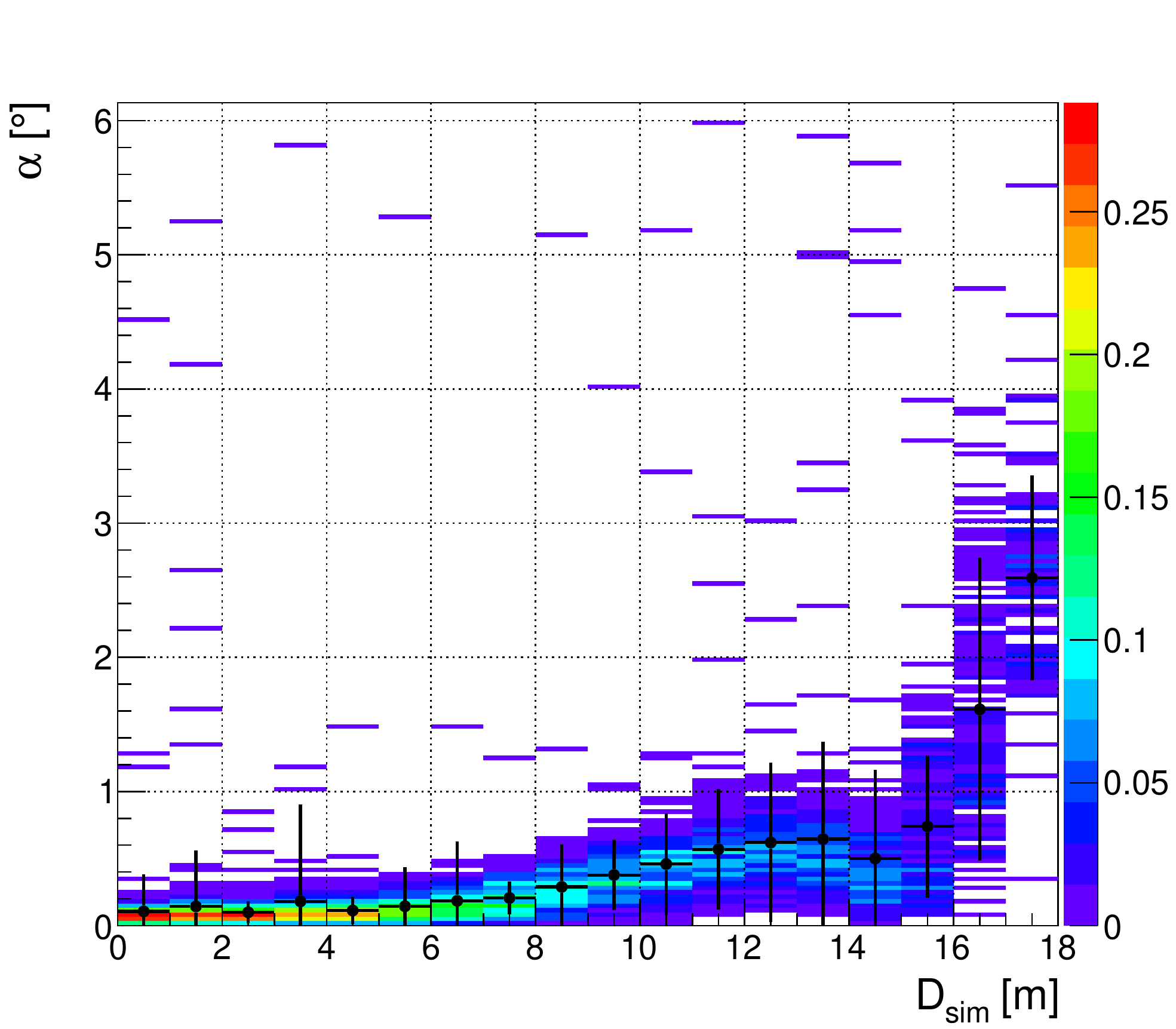}}%
\hfill
\subcaptionbox*{LPMT+SPMT $\Delta D$}{\includegraphics[width=0.40\textwidth]{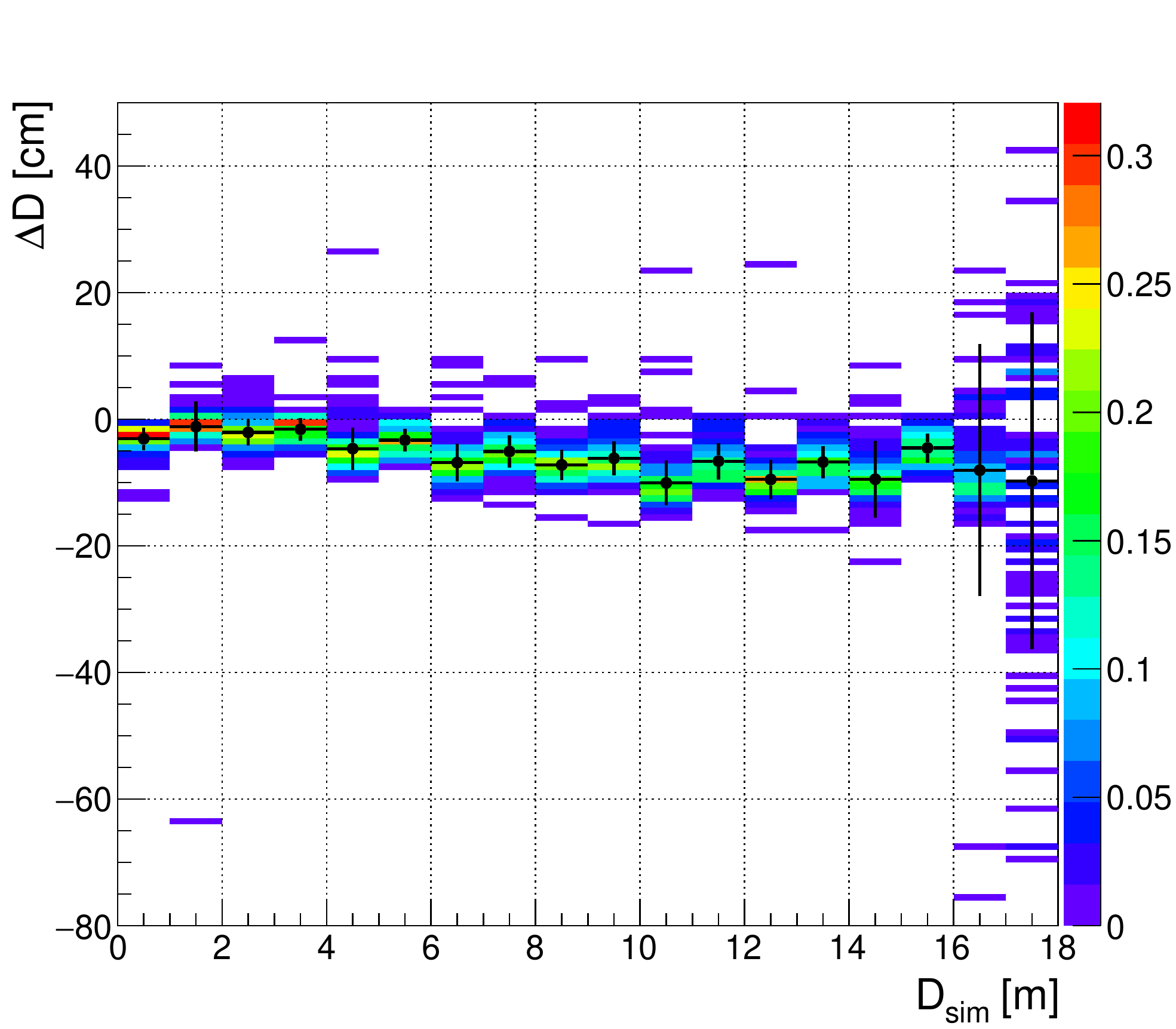}}%
\subcaptionbox*{LPMT+SPMT $\alpha$}{\includegraphics[width=0.40\textwidth]{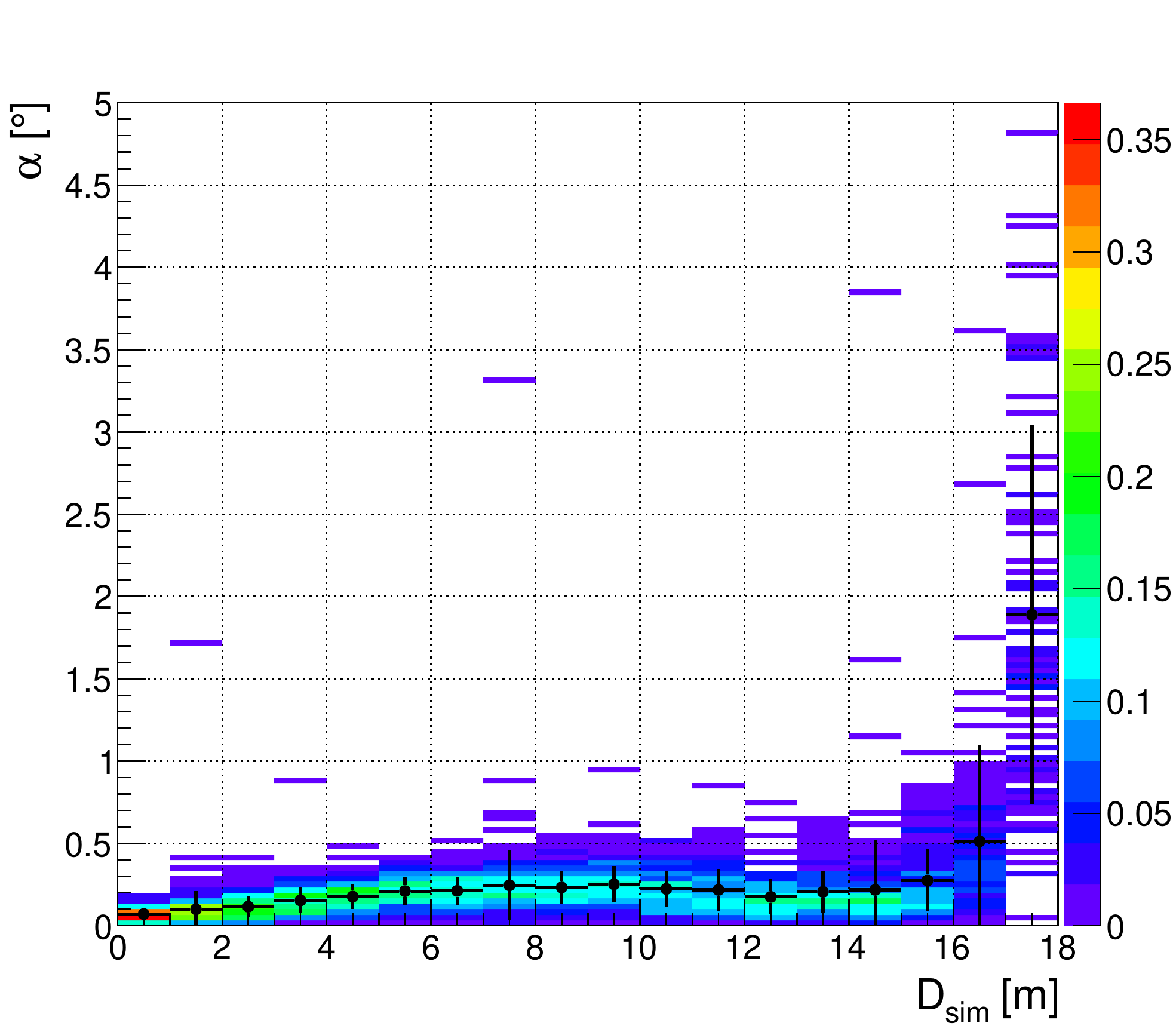}}%
\caption{\label{fig:performance_noelec} Reconstruction results for a sample of 5900 simulated muon tracks with smeared hit times according to the PMT's TTS. The first row shows results when using only the LPMT's array. In the second row only the SPMT's system was used and the third row presents the combination of both systems into one fit. In this case, the deviation in distance from center $\Delta D$ shows a small mean bias of less than \unit[10]{cm}. The bias in angular reconstruction is better than $\unit[0.5]{^\circ}$ for the largest part of the detector. On the very edge of the sphere, the reconstruction performance declines because the muon travels only a short distance through the LS.}
\end{figure}
\begin{figure}[htbp]
\centering
\includegraphics[width=\textwidth]{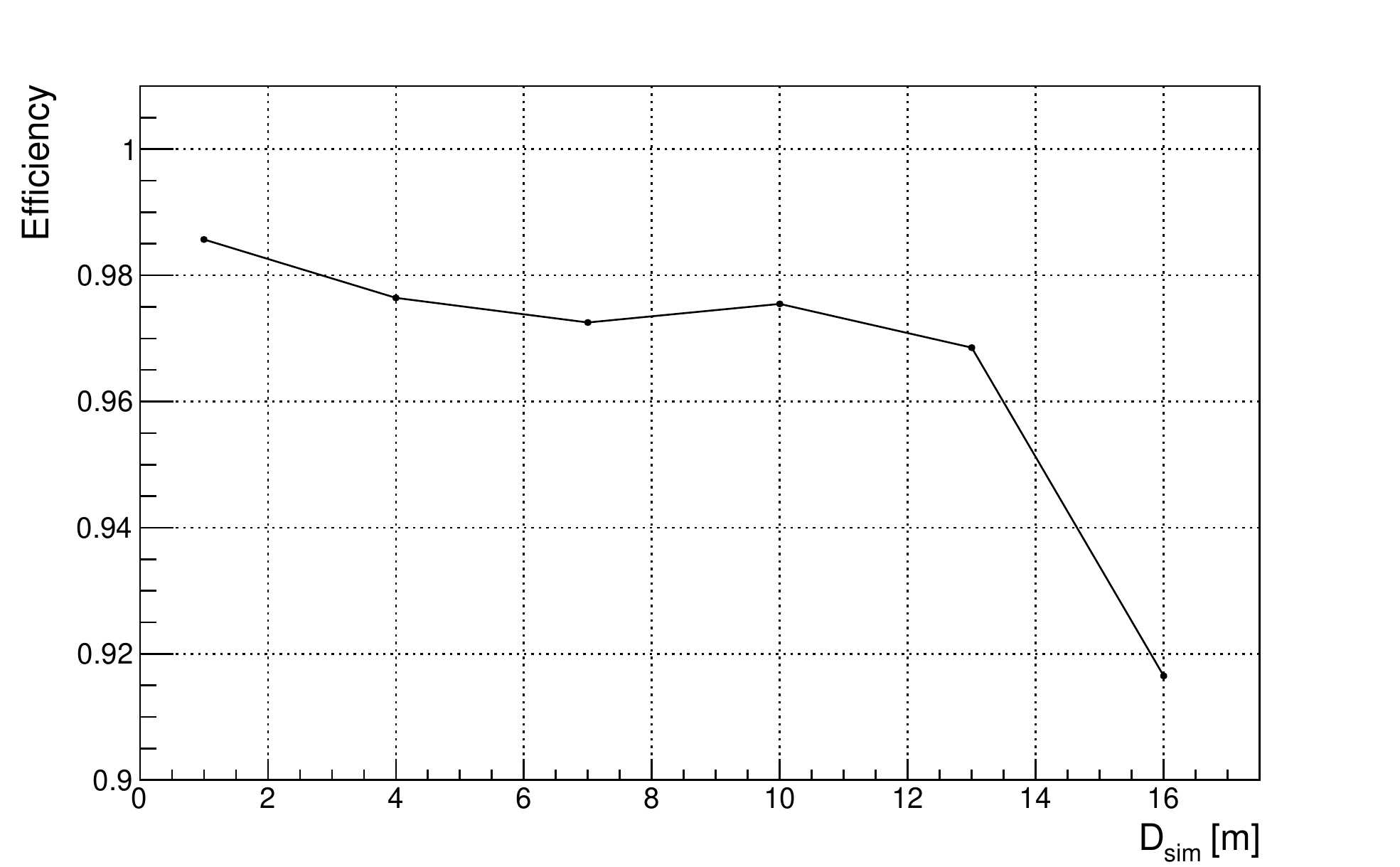}
\caption{\label{fig:efficiency} The reconstruction efficiency when using the combined system of LPMT and SPMT.}
\end{figure}

Figure \ref{fig:performance_noelec} shows the algorithm's performance for the different PMT systems. In this mode, the PMT's quantum efficiency and collection efficiency is simulated and the hit times are smeared by a Gaussian distribution according to their respective transit time spread as explained in section \ref{sec:testprocedure}. The first row shows the performance that can be achieved with only the LPMT system. There is a strong increase in bias for $\Delta D$ for increasing $D$. This bias development can be explained with taking the amount of collected light into account. When collecting a lot of light within a short period of time, the extracted first hit time tends to be systematically earlier. PMTs that have an earlier hit time pull the track towards themselves. This effect increases towards the detector edge, when the tracks comes closer to the PMTs and they collect even more light. The second row displays the results for the isolated SPMT system. The bias develops in the opposite direction as with the LPMTs, but stays below \unit[20]{cm}. Nevertheless, it has a larger spread around the mean values. Since the SPMTs are much smaller and also slightly shadowed by the LPMT, they do not suffer from the effect of pulling tracks towards the edge, as explained above. In this case the bias only significantly increases for values of $D>\unit[15]{m}$, where the influence of  refraction on the acrylic sphere increases. In the bottom row, both PMT systems, \unit[20]{inch} and \unit[3]{inch}, are used in a combined fit. The reconstruction runs stable with a small mean bias in $\Delta D$ of less than \unit[10]{cm} for most of the tracks. Both PMT systems isolated exhibit a bias that pulls the tracks in opposite directions when increasing the track's true distance from center. In the combined fit, the effects of the complementary systems cancel out. In the same region, also the angular bias stays below $\unit[0.5]{^\circ}$. Only the last bin for tracks with $D\geq \unit[17]{m}$ features a worse resolution due to the very short tracklength.
The tracking efficiency is calculated by the ratio of well reconstructed tracks to all tracks. A track is considered well reconstructed if the deviation of each parameter is less then five times their standard deviation. The overall efficiency is better than 96\%. As shown in figure \ref{fig:efficiency}, reconstruction in the inner \unit[13]{m} has an efficiency around 97\%, while for edge tracks the efficiency declines to 91\%.

\begin{figure}[htbp]
\centering
\subcaptionbox*{}{\includegraphics[width=0.50\textwidth]{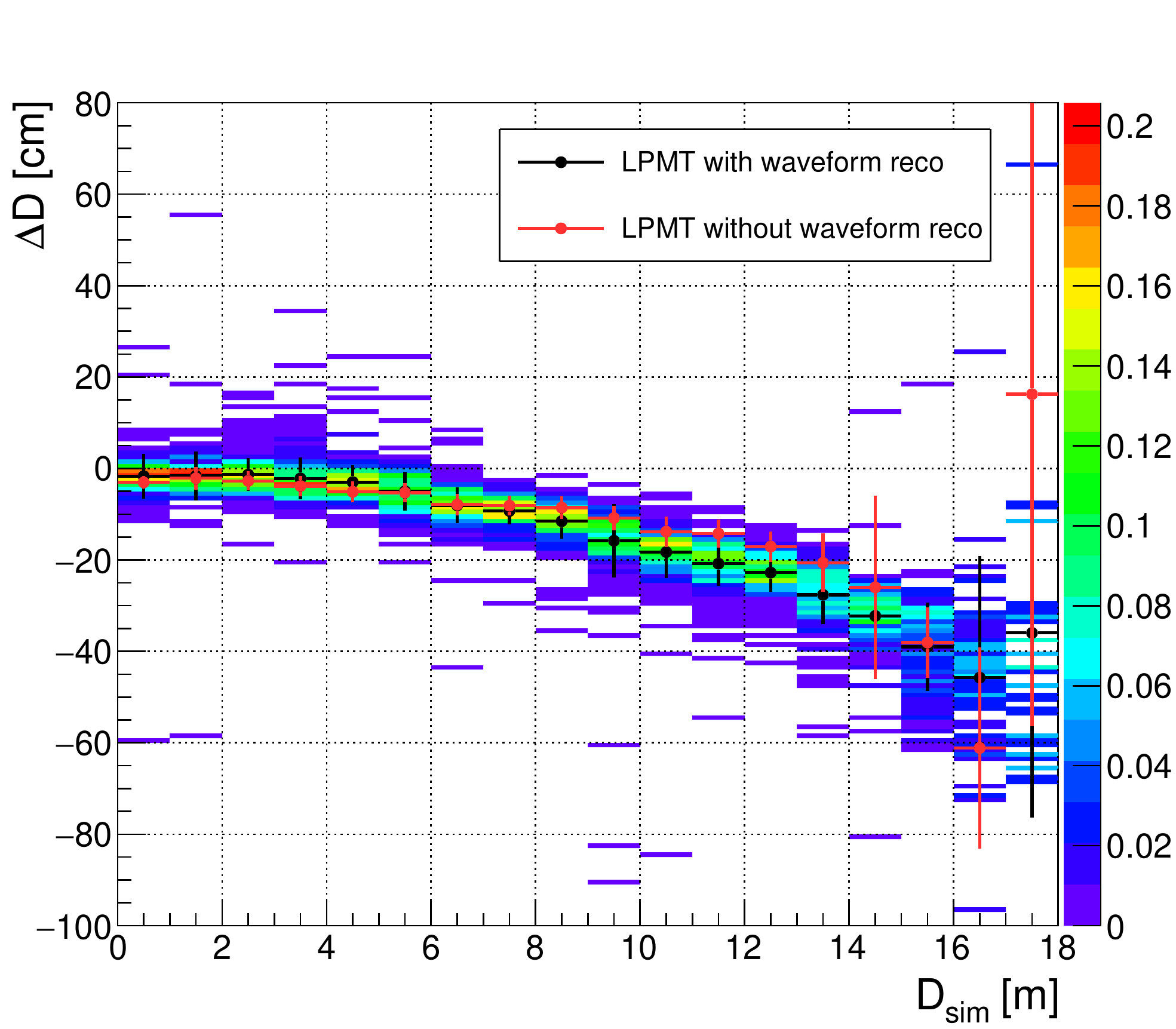}}%
\hfill
\subcaptionbox*{}{\includegraphics[width=0.50\textwidth]{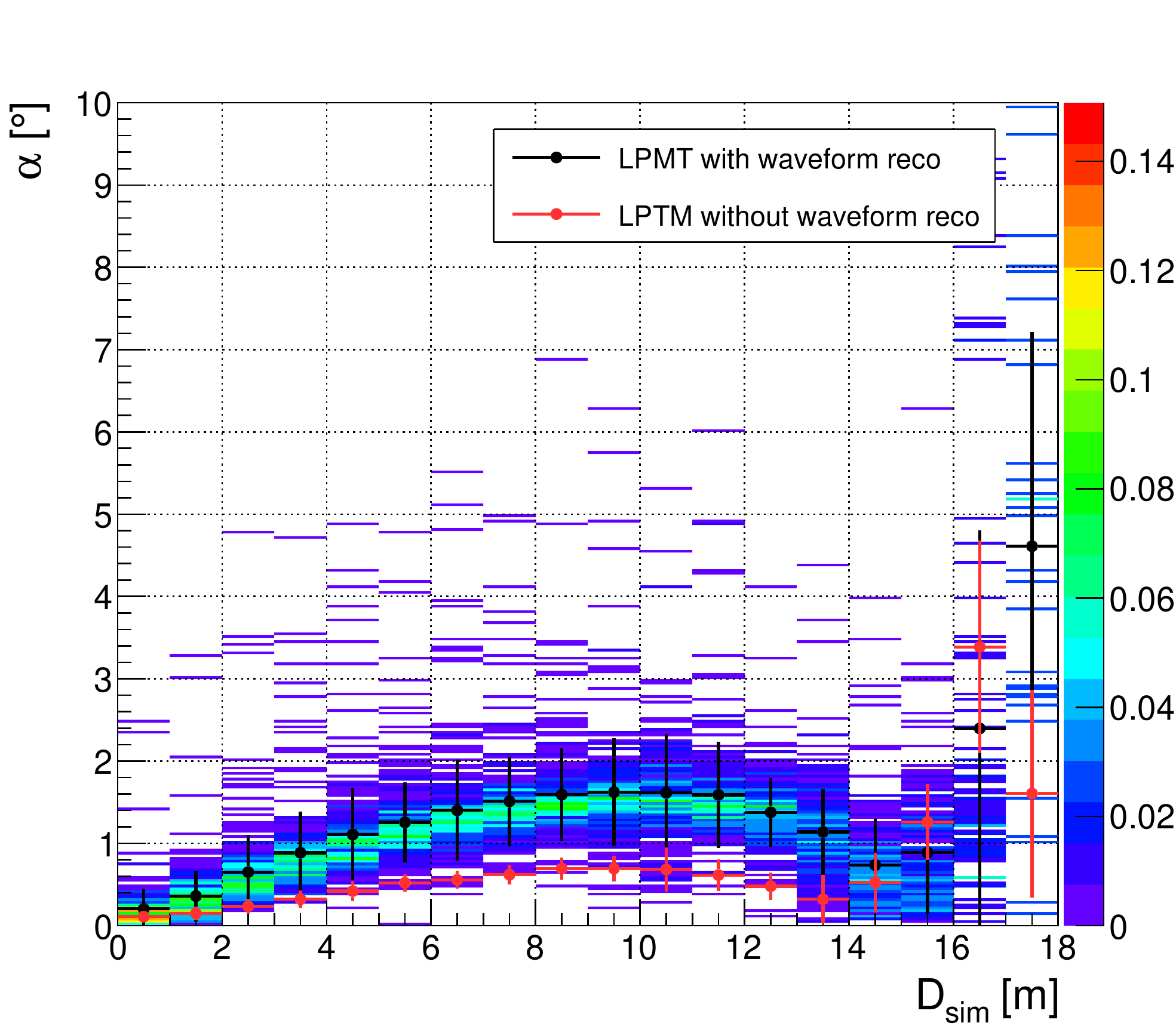}}%
\caption{\label{fig:performance_elec} Reconstruction results for a sample of 5940 simulated muon tracks with waveform reconstruction. Here the deviation in the muon track's distance from the detector center $\Delta D$ has an increasing mean bias of less than \unit[50]{cm}. The bias in angular reconstruction is better than $\unit[2]{^\circ}$ for the largest part of the detector. Tracks at the very edge of the detector have a significantly larger spread similar to the case without waveform reconstruction.}
\end{figure}

In a second step, the method was verified when using the additional step of the waveform reconstruction. The results are shown in figure \ref{fig:performance_elec}.
In this mode, only for the large \unit[20]{inch} PMT the waveform reconstruction is done explicitly. The SPMT system is not designed to output waveforms, but direct counts for charges and hit times. Due to that, there is no need for a waveform reconstruction for SPMT. The largest uncertainty of this system is described by smearing the first hit times according to their TTS. However, for the LPMT system more distortion is introduced to the signal through another reconstruction step from waveforms to first hit times and charges. Accordingly, the performance is worsened. With increasing distance $D$ from center, the bias in $\Delta D$ increases as well up to the mean value of \unit[40]{cm}. Also the angular bias is increased to value up to $\unit[1.5]{^\circ}$. Both modes have in common that the last bins for $D\geq \unit[17]{m}$ show no reliable results for corner-clipping tracks. For comparison the LPMT profile without waveform reconstruction is shown in red.

This performance can compare to established experiments, which are currently taking data. Borexino reports a muon tracking algorithm with a lateral resolution between \unit[35-50]{cm} and an angular resolution of 3-$\unit[5]{^\circ}$ \cite{borexMuon:2011}. Around the detector center, Borexino can reach a spatial resolution of \unit[30]{cm}.
The high precision muon tracking in Double Chooz can even reach a spatial resolution of \unit[4]{cm} in each transverse direction for tracks close to the center of the detector \cite{Abe:2014yda}, while the shortest tracks are reconstructed with a spatial resolution of 10-\unit[15]{cm}.
Within this context, the introduced cone algorithm promises an improvement in resolution compared to Borexino and being on about the same level as the muon tracking in Double Chooz. Nevertheless, a direct comparison by those numbers is not straightforward, since several other aspects like PMT coverage as well as detector geometry and size have to be taken into account.
In order to compare JUNO to other multi-kiloton LS detectors, there are only Monte Carlo based studies available. The studies for the LENA detector \cite{WURM2012685} included a muon tracking algorithm which is also based on PMT hit times \cite{HELL2015}. This reconstruction was tested on the simulation of contained lower energy muons between \unit[0.2]{GeV} and \unit[1]{GeV}. The reached resolution on the vertex is better than \unit[10]{cm}, while the angular resolution could reach $\unit[1.7]{^\circ}$. Although the test sample features muons with shorter track lengths, it is also compatible with the results of \unit[5]{cm} spatial resolution and $\unit[0.3]{^\circ}$ angular resolution obtained here for high energy through-going muons.

\section{Deadtime estimation}
\label{sec:deadtime}
As explained in section \ref{sec:intro}, muons are responsible for a substantial amount of detector dead time due to the need to veto cosmogenic isotopes. With a partial veto of a volume around the muon track, this loss of exposure can be reduced significantly.
The effect of the muon veto on detector deadtime is quantified with help of a toy Monte Carlo simulation. As input, we use a simulated muon flux of 75000 tracks generated as presented in \cite{yb:2015}. This spectrum is considering the energy and angular distribution of muons arriving at the central detector. The baseline veto strategy is a cylindrical volume with $r_{v} = \unit[3]{m}$ for \unit[1.2]{s} after a muon along its track through the CD. This veto can reduce the cosmogenic background by 98\% \cite{yb:2015}. In accordance with a muon rate of $\unit[3]{s^{-1}}$, the vetoed and sensitive detector volume is numerically determined whenever a new muon enters the CD or an existing veto cylinder is released. The volume integration is done with a grid of $\sim$4.2$\cdot 10^9$ points, which corresponds to a volume of $\unit[5.5]{cm^3}$ per point. This approach takes into account that several veto cylinders are present in the detector at the same time and that they can overlap.
In order to quantify the reconstruction efficiency, the cylinder radius $r_{v} = \unit[3]{m}$ is increased in proportion to the biases in $\Delta D$ and $\alpha$ given above. The resulting effective veto radius is estimated as
\begin{equation}
r_{v,\rm eff} = r_v + \Delta D + \sin(\alpha) l,
\end{equation}
with $l = \sqrt{R_{LS}^2 - D^2}$ being the half track length in LS.
The results are summarized in table \ref{tab:deadtime}. The loss of 14\% exposure with perfect tracking is unavoidable with the applied veto strategy and acts as a benchmark value for the developed reconstruction algorithm. With full simulation and waveform reconstruction the loss of exposure increases only to 18\% in total when using only the LPMTs. Thus, the imperfection of the reconstruction algorithm adds only 4\% of exposure-loss. According to the improved performance of the combined LPMT+SPMT system the loss of 4\% is a conservative estimate.
\begin{table}[htb]
	\begin{center}
	\ra{1.2}
	\caption{\label{tab:deadtime} Summary of deadtime estimation in terms of exposure ratio. The efficiency of 86\% for a perfect tracking is in accordance with the reported muon veto efficiency for IBD events \cite{yb:2015}. The results for the reconstruction with the cone model are seperated as explained in section \ref{sec:performance}. \label{tab:deadtimeTable}}
	\begin{tabular}{@{} lr @{}}\toprule
	Veto strategy & Exposure ratio \\ \midrule
	No veto & 100\% \\
	Perfect tracking & 86\% \\
	ConeReco LPMT+SPMT & 85\% \\ 
	ConeReco LPMT with waveform reconstruction & 82\% \\ \bottomrule
	\end{tabular}
	\end{center}	
\end{table}

\section{Conclusion}
\label{sec:conclusion}
A sophisticated implementation of the fastest-light approach for muon tracking was developed and tested for JUNO. It is based on the universally applicable geometrical model of fastest light propagation and its intersection with an arbitrary shape of a PMT array. We have shown that in the largest part of the detector, the algorithm can reconstruct muon tracks with a bias of less than \unit[45]{cm} in $D$ and $\unit[1.5]{^\circ}$ in direction. A resolution better than \unit[5]{cm} in $D$ and $\unit[0.3]{^\circ}$ in direction can be reported in this region. The performance closer to the detector's edge could benefit from including effects of refraction on the acrylic sphere into the cone model. Given the proposed veto scheme \cite{yb:2015}, a reduction of only 4\% in exposure is expected due to the waveform and track reconstruction itself for a realistic muon flux. Without the effects of waveform reconstruction the exposure loss introduced by the reconstruction is only 1\% with both LPMTs and SPMTs combined. In a further step, the top tracker of JUNO could be included to measure a precisely tracked muon sample. This could be used to calibrate muon reconstruction algorithms to decrease the bias. 
The geometrical approach models the scintillation light front that develops in any liquid scintillator. It fits the timing signal of the PMTs to the intersection of two geometrical shapes --- the light cone and an arbitrarily shaped PMT array. Therefore it can be applied also in other liquid scintillator detectors.

\acknowledgments
This work was funded through the HGF Recruitment Initiative and it was also supported by Deutsche Forschungsgemeinschaft DFG, Forschergruppe JUNO (FOR 2319). In addition, we acknowledge the strong support by the JUNO collaboration, which provided us with the software framework and detector simulation to test the described reconstruction with the JUNO detector. Furthermore we would like to thank Sebastian Lorenz, Bj{\"o}rn Wonsak, and Michael Wurm for fruitful discussions along the development and analysis of the method.

\bibliographystyle{JHEP}
\bibliography{./bib/theBib}

\end{document}